\documentclass[12pt,preprint]{aastex}

\usepackage{natbib}
\usepackage{epsfig}
\newcommand\cm{{\,\rm cm}}
\newcommand\pcc{{\,\rm cm}^{-3}}
\newcommand\K{{\;\rm K}}
\newcommand\vth{v_{\rm th}}
\newcommand\yr{{\;\rm yr}}
\newcommand\g{{\;\rm g}}
\newcommand\Gyr{{\;\rm Gyr}}
\newcommand\Myr{{\;\rm Myr}}
\newcommand\Msun{{\;\rm\,M_\odot}}
\newcommand\kms{{\;\rm km\; s^{-1}}}
\newcommand\pc{{\;\rm\,pc}}
\newcommand\kpc{{\;\rm kpc}}
\newcommand\simgt{\lower.5ex\hbox{$\; \buildrel > \over \sim \;$}}
\newcommand\simlt{\lower.5ex\hbox{$\; \buildrel < \over \sim \;$}}
\newcommand\epsff{{\, \varepsilon_{\rm ff}}}
\newcommand\nthr{{\, n_{\rm thr}}}
\newcommand\tff{{\, t_{\rm ff}}}
\newcommand\cs{{\, c_{s}}}
\newcommand\sigsfr{\Sigma_{\rm SFR}}

\newcommand\mgmc{M_{\rm GMC}}
\newcommand\epsgmc{\varepsilon_{\rm GMC}}

\newcommand\tsfgas{t_{\rm SF,gas}}
\newcommand\vel{\mathbf{v}}

\slugcomment{Accepted for publication in the ApJ}

\shorttitle{Star-Forming Disks}
\shortauthors{Ostriker \& Shetty}

\begin{document}


\title{Maximally Star-Forming Galactic Disks I. Starburst Regulation Via
  Feedback-Driven Turbulence}


\author{Eve C. Ostriker\altaffilmark{1} and Rahul Shetty\altaffilmark{2}}
\altaffiltext{1}{Department of Astronomy, University of Maryland, College Park,
  MD 20742} 
\altaffiltext{2}{Zentrum f\"ur Astronomie der Universit\"at Heidelberg, 
Institut f\"ur Theoretische Astrophysik, Albert-Ueberle-Str. 2, 69120 
Heidelberg, Germany}
\email{ostriker@astro.umd.edu,rshetty@ita.uni-heidelberg.de}

\begin{abstract}
Star formation rates in the centers of disk galaxies often vastly
exceed those at larger radii, whether measured by the surface density
of star formation $\sigsfr$, by the star formation rate per unit gas
mass, $\sigsfr/\Sigma$, or even by total output.  In this paper, we
investigate the idea that central starbursts are self-regulated
systems, in which the momentum flux injected to the interstellar
medium (ISM) by star formation balances the gravitational force
confining the ISM gas in the disk.  For most starbursts, supernovae
are the largest contributor to the momentum flux, and turbulence
provides the main pressure support for the predominantly-molecular
ISM.  If the momentum feedback per stellar mass formed is $p_*/m_*\sim
3000 \kms$, the predicted star formation rate is $\sigsfr \sim 2 \pi G
\Sigma^2 m_*/p_*\sim 0.1 \Msun \kpc^{-2} \yr^{-1} (\Sigma/100 \Msun
\pc^{-2})^2$ in regions where gas dominates the vertical gravity.  We
compare this prediction with numerical simulations of
vertically-resolved disks that model star formation including
feedback, finding good agreement for gas surface densities in the
range $\Sigma \sim 10^2 -10^3 \Msun \pc^{-2}$.  We also compare to a
compilation of star formation rates and gas contents from local and
high-redshift galaxies (both mergers and normal galaxies), finding good
agreement provided that the conversion factor $X_{CO}$ from integrated
CO emission to $H_2$ surface density decreases weakly as $\Sigma$ and
$\sigsfr$ increase.  Star formation rates in dense, turbulent gas are
also expected to depend on the gravitational free-fall time at the
corresponding mean ISM density $\rho_0$; if the star formation
efficiency per free-fall time is $\epsff(\rho_0) \sim 0.01$, the
turbulent velocity dispersion driven by feedback is expected to be
$v_z = 0.4 \epsff(\rho_0) p_*/m_*\sim 10\kms$, relatively independent
of $\Sigma$ or $\sigsfr$.  Turbulence-regulated starbursts (controlled
by kinetic momentum feedback) are part of the larger scheme of
self-regulation; primarily-atomic low-$\Sigma$ outer disks may have
star formation regulated by UV heating feedback, whereas regions at
extremely high $\Sigma$ may be regulated by feedback of stellar radiation that
is reprocessed into trapped IR.

\end{abstract}


\section{Introduction}

\begin{figure}[t]
\epsscale{.8}
\plotone{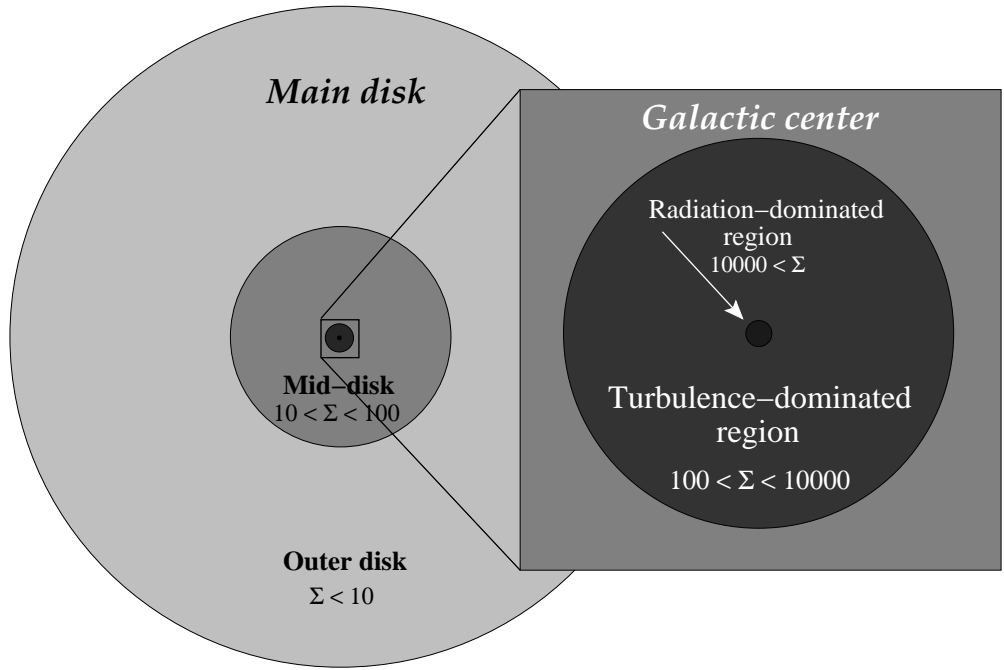}
\caption{Schematic indicating the geography of galactic disk star
  formation; ranges for gaseous surface density $\Sigma$ (in $\Msun
  \pc^{-2}$) in the various regions are approximate.  Regulation of
  star formation in the main disk has recently been discussed by
  \citet{Ostriker2010}.  In this paper, we analyze self-regulation of
  star formation in the turbulence-dominated region within the galactic
  center, for starburst disks.
\label{fig1}}
\end{figure}

Averaged over large scales in disk galaxies, the relationship between
the mean surface density of star formation, $\sigsfr$, and the mean
gaseous surface density, $\Sigma$, is observed to be superlinear, both
in the local Universe and at higher redshift
(e.g. \citealt{Kennicutt98,2010MNRAS.407.2091G}). This behavior
reflects the increased efficiency (or shorter timescale) of star
formation under conditions of higher mean gas density, which is
correlated with higher $\Sigma$.  Starburst regions within the central
$\sim \kpc$ of galaxies, commonly observed as dust-enshrouded LIRGs
and ULIRGs in which the ISM is predominantly
molecular
\citep{1996ARA&A..34..749S,1998ApJ...498..579G,Kennicutt98AR,2005ARA&A..43..677S},
represent the extreme of this behavior, with $\Sigma \sim 10^2 -10^4
\Msun \pc^{-2}$ and the star formation (or gas conversion) timescale
$\tsfgas\equiv \Sigma/\sigsfr$ a factor of 10 or more below $\tsfgas$ in
lower-$\Sigma$ regions of galaxies.  Although earlier observations
based on global averages suggested a simple power-law relationship
between $\Sigma$ and $\sigsfr$, these global
relations have considerable scatter about the mean.  More recently,
evidence has emerged that the star formation regime within galactic
centers at very high $\Sigma \simgt 100 \Msun \pc^{-2}$ is
distinct from the star formation regime that prevails in the main
disks of galaxies at lower $\Sigma \simlt 100 \Msun \pc^{-2}$.  Within
the main disk, another change in the star formation regime appears to
occur from ``mid-disk'' to ``outer-disk.'' Typically, both galactic
center and mid-disk regions are primarily molecular, whereas outer
disks are primarily atomic.  Figure (\ref{fig1}) 
provides a schematic, dividing the disk into different star-forming regimes.

In main-disk regions where $\Sigma \simlt 100 \Msun \pc^{-2}$,
high-resolution observations have found that $\sigsfr$ is proportional
to $\Sigma_{\rm mol}$, the mean surface density of molecular gas, with
a timescale $t_{\rm SF, mol} \equiv \Sigma_{\rm mol}/\sigsfr \approx
2\Gyr$ \citep{Bigiel08,2009ApJ...704..842B}.  The value
$\Sigma \sim 100 \Msun \pc^{-2}$ that defines the upper limit for
``main disks'' is characteristic of the surface density $\Sigma_{\rm
  GMC}$ of individual giant molecular clouds (GMCs), as observed in
the Milky Way and Local Group \citep{2007prpl.conf...81B,She08,Bol08}.
The relation $\sigsfr \propto \Sigma_{\rm mol}$ for $\Sigma \simlt
\Sigma_{\rm GMC}$ is consistent with the idea that spatially-isolated
GMCs have an approximately constant specific star formation rate; this
rate, and the value of $\Sigma_{\rm GMC}$, may be controlled by
internal feedback processes within GMCs
(e.g. \citealt{2006ApJ...653..361K}), although this is not yet completely 
understood. Over a typical GMC lifetime of
$\sim 20 \Myr$ \citep{2007prpl.conf...81B}, $\sim 1\%$ of the gas must
be converted to stars in order to yield $t_{\rm SF, mol} \approx 2\Gyr$ within main-disk regions.  Although there are uncertainties
arising from limited resolution and calibration of the CO-to-H$_2$
conversion (see Section 4), the relationship between $\Sigma_{\rm
  mol}$ and $\sigsfr$ in galactic center regions with $\Sigma \simgt
100 \Msun \pc^{-2}$ appears considerably steeper than linear.
Presumably, this is because the molecular gas in galactic center
regions with $\Sigma \simgt 100 \Msun \pc^{-2}$ is at higher mean
(volume) density than the molecular gas in main-disk GMCs having
``standard'' surface density $\Sigma_{\rm GMC} \sim 100 \Msun
\pc^{-2}$.

Molecular gas dominates mid-disk regions, but is a small fraction of
the total in outer disks, where $\Sigma$ and the disk's stellar
density $\rho_*$ are lower
\citep{2002ApJ...569..157W,BR04,BR06,Ler08}.  \citet{Ostriker2010}
have recently advanced a theory to explain the star-forming behavior
and balance of phases in the main regions of disks at $\Sigma \simlt
100 \Msun \pc^{-2}$, based on the requirement that the diffuse
\ion{H}{1} gas must be in both thermal and dynamical equilibrium.  In
this theory, star-forming gravitationally-bound clouds are assumed
to have internal pressures much larger than their surroundings, 
consistent with local GMC observations.  Thus, they are 
effectively isolated, and over their lifetimes 
are assumed to maintain a mean internal 
specific star formation rate consistent with the empirical main-disk 
value $t_{\rm SF, mol}\approx 2\Gyr$.  In equilibrium,
individual gravitationally-bound clouds form and disperse at equal
rates, with neutral ISM gas cycling between the diffuse and
gravitationally-bound components.  In outer-disk regions (typically
$\Sigma \simlt 10 \Msun \pc^{-2}$, although this varies with
$\rho_*$), the equilibrium  
fraction of gas in star-forming bound clouds is just
enough that the heating provided by the resulting stellar UV balances
the cooling of the surrounding diffuse \ion{H}{1}; here, $\sigsfr
\propto \Sigma \sqrt{\rho_*}$.  In mid-disk regions (typically $10
\simlt \Sigma \simlt 100 \Msun \pc^{-2}$), the interstellar medium
becomes predominantly molecular (by mass) because the heating provided
by star formation would be insufficient to balance cooling in
high-density warm atomic gas (which is compressed by the vertical disk
gravity); here $\sigsfr \propto \Sigma_{\rm mol} \propto \Sigma$.  The
model of \citet{Ostriker2010} assumes that most of the volume is
filled by non-star-forming atomic gas, and thus is inapplicable to the
starburst regime in galactic centers where $\Sigma \simgt 100 \Msun
\pc^{-2}$ and gas is mostly molecular.

Where $\Sigma$ is very high, the conditions throughout the ISM
resemble those in the interior of locally-observed GMCs, but are even
more extreme in terms of the mean densities, the abundance of very
high density gas, and the ratio of the turbulent velocity dispersion
$v_{\rm turb}$ to the thermal speed $\vth$
(e.g. \citealt{1991ApJ...366L...5S,1993ApJ...414L..13D,1998ApJ...507..615D,
  2004ApJ...606..271G}).  Supersonic turbulence is known to be a
dominant process controlling star formation regulation within GMCs
\citep{2004RvMP...76..125M,2007ARA&A..45..565M}, so it is expected to
be important in starburst regions as well.  Main-disk GMCs are
relatively isolated entities, in the sense that their internal
dynamical timescales are short compared to the orbital, epicyclic and
vertical oscillation times in the galactic potential ($\Omega^{-1}$,
$\kappa^{-1}$, and $\nu^{-1}$, respectively).  In galactic center
regions, however, the gravitational potential is deeper, such that
$\Omega$, $\kappa$, and $\nu$ are all substantially higher, with
timescales $\simlt 10 \Myr$.  Thus, the galactic environment may also
significantly impact the development of star formation.

Turbulence in the ISM both inhibits and encourages star formation.
For regions large compared to the energy (and momentum) injection
scale, turbulence provides an effective pressure that opposes gravity.
In the ISM, this helps to regulate the mean density averaged over the
disk thickness $\bar\rho$, and therefore the large-scale
self-gravitational timescale $\langle t_{\rm grav} \rangle \propto
\bar\rho^{-1/2}\propto (H/\Sigma)^{1/2}$.  Here, the gas disk
semi-thickness $H$ varies with turbulent velocity dispersion as
$H\propto \sigma_z$ or $\sigma_z^2$ depending on whether stellar or
gas gravity dominates the vertical potential gradients.  In either
case, larger turbulent velocities raise $H$ and $\langle t_{\rm
  grav}\rangle $ for the ISM disk, while larger stellar and gas
densities lower $H$ and $\langle t_{\rm grav}\rangle $.  On scales
smaller than the turbulent injection, supersonic compression creates
local overdense regions $\rho_{\rm local} \gg \bar\rho $ that
potentially can collapse more rapidly than the larger-scale regions
containing them since $t_{\rm grav,local}/\langle t_{\rm grav} \rangle
\propto (\rho_{\rm local}/\bar\rho)^{-1/2}\ll 1$.  However, turbulent
rarefactions and shear also destroy overdensities.  Because $t_{\rm
  grav,local}$ depends on density, turbulence will disperse the
moderate-density structures before self-gravity can concentrate them
further, while allowing the highest-density regions (amounting to a
small fraction of the mass) to collapse and form stars.  Theoretical
arguments suggest that the fraction of mass in turbulent systems
that collapses to form stars will be relatively insensitive to the
velocity dispersion (\citealt{KM05}; see also Section 5), so that the
overall star formation rate is primarily governed by the gravitational
time on large scales, $\langle t_{\rm grav}\rangle $.  Since massive
stars energize the ISM, raising the velocity dispersion and $\langle
t_{\rm grav}\rangle$, the associated feedback loop may allow star
formation rates to be self-regulated.

In this paper, we investigate the process of self-regulation via
turbulent driving, for application to understanding what controls star
formation rates in starburst disks.  For molecule-dominated regions,
cooling times are short, so it is the injection of momentum, rather
than energy, that is essential in defining the dynamical state of the
disk.  Thus, we begin in Section 2 by considering the implications of
maintaining force balance in the vertical direction, which imposes a
requirement on the input momentum flux associated with star formation
feedback.  We show that for disks where supernovae dominate the
feedback, the star formation rate is expected to vary as $\sigsfr
\propto \Sigma^2$.  At very high surface density in optically-thick disks, 
radiation pressure
can exceed the turbulent pressure driven by supernovae, which leads to
$\sigsfr \propto \Sigma$.  In Section 3, we compare our prediction for
$\sigsfr$ vs. $\Sigma$ in turbulence-dominated disks to the results of
numerical simulations, for which feedback is explicitly implemented in
local vertically-resolved models of self-gravitating, rotating disks.
Section 4 compares the prediction for $\sigsfr$ vs. $\Sigma$ to
observations of galaxies in the regime where the turbulence-dominated
model is expected to apply.  Both the numerical simulations and the
observations are consistent with the simple analytic
model.  In Section 5, we connect to models relating star formation
rates to $\langle t_{\rm grav}\rangle$, and discuss how the mean
internal properties (including velocity dispersion and disk thickness)
of turbulence-dominated, feedback-regulated disks are expected to
depend on $\Sigma$ and the parameters that characterize the star
formation feedback.  We conclude in Section 6 with a brief summary and 
discussion.

\section{A simple model for momentum-controlled, self-regulated disk 
star formation}

Consider a region of a galactic disk with gas surface density
$\Sigma$, residing within a bulge of uniform density $\rho_b$, such
that the gas-disk gravitational potential is $\Phi_g$ and the bulge
potential is $\Phi_b$.  We assume that the gas disk's internal
structure is similar to that of a GMC, in that it is highly
inhomogeneous (due to supersonic turbulence) but most of the gas
concentrations are transient.  The disk is thus taken to consist
primarily of ``diffuse'' gas, in the sense that only a small fraction
of the mass at any time lies in bound clumps
having gravitational potential large compared to the mean
midplane value in the disk. With an effectively plane-parallel
distribution of mean gas density, the mean gravitational potential depends
only on the distance from the midplane $z$. The time-averaged,
horizontally-averaged vertical momentum equation
(e.g. \citealt{1990ApJ...365..544B,PO07,KO09b,Ostriker2010}) may then
be vertically integrated, yielding the result that the total momentum
flux through the disk must be equal to the total vertical weight of
the overlying gas.

At the midplane, the weight includes a self-gravitational term,
\begin{equation}
\int_0^\infty \rho \frac{d \Phi_g}{dz} dz
=\frac{1}{8\pi G} \int_0^\infty 
\frac{d\left(\frac{d \Phi_g  }{d z  }\right)^2 }
{d z  } dz 
=\frac {\pi G \Sigma^2  } { 2 },
\label{gas_weight}
\end{equation}
where we have used $\rho = (4 \pi G )^{-1} d^2 \Phi_g/dz^2$ and 
$\vert d \Phi_g/dz\vert_\infty = 2 \pi G \Sigma$
for a slab; and a term arising from the external bulge 
potential\footnote{Note that the expression (\ref{bulgeweight})
  for a stellar bulge differs from the corresponding equation for a
  stellar disk in \citet{Ostriker2010} by a factor $1/3$.},
\begin{equation}
\int_0^\infty \rho \frac{d \Phi_b}{dz} dz
=\frac{4\pi G \rho_b  }{3  } \int_0^\infty \rho z dz 
\equiv\frac{2\pi \zeta_d G \rho_b \Sigma^2} {3 \rho_0  }.
\label{bulgeweight}
\end{equation}
Here, $\rho(z)$ is the mean (area-weighted) density of the gas at
height $z$, and $\rho_0$ is the midplane value.  The value of the
numerical coefficient $\zeta_d$ is insensitive to the exact vertical
distribution of the gas, equaling $1/\pi$ for a Gaussian (when
$\Phi_b$ dominates) and $\ln(2)/2$ for a $\rm sech^2$ distribution
(when $\Phi_g$ dominates); $\zeta_d \approx 0.33$ within $5\%$ for
these distributions.  Note that the self-gravitational weight is
independent of the midplane gas density, whereas the weight in the
external bulge potential is proportional to
$\rho_b/\rho_0$.

The difference in momentum flux between the disk midplane and the height 
$z_{\rm max}$ at which $\rho \rightarrow 0$ includes thermal and turbulent 
terms, $P_{\rm th}=\rho_0 \langle \vth^2\rangle $ 
and $P_{\rm turb}=\rho_0 \langle v_z^2 \rangle $, where 
$\langle v_z^2\rangle $ 
is equal to $1/3$ of the total turbulent velocity variance $v_{\rm turb}^2$ 
if it is isotropic. 
The thermal and turbulent terms can be combined as a single midplane 
kinetic pressure
$\rho_0 \sigma_z^2$, where $\sigma_z$ is the observed velocity dispersion for
a face-on disk.
Note that $P_{\rm th}$ corresponds to a volume-weighted mean pressure, and 
$\langle \vth^2\rangle $ 
corresponds to a mass-weighted average of thermal
speeds of different components:
if $M_{total}=\sum_i M_i $ and 
$V_{total}= \sum_i V_i $ for mass and volume 
with $\rho_i=M_i/V_i=P_{\rm th,i}/v_{\rm th,i}^2$, then 
$P_{\rm th}\equiv \sum_i V_i P_{\rm th,i}/V_{total}=
\sum_i M_i v_{\rm th,i}^2/V_{total}=
\rho_0\sum_i (M_i/M_{total}) v_{\rm th,i}^2\equiv
\rho_0 \langle v_{\rm th}^2 \rangle$.
For molecule-dominated regions, the RMS thermal speed 
$\langle \vth \rangle^{1/2} \simlt 1 \kms$ if 
$T\simlt 100 \K$, whereas observed velocity dispersions are at least several
$\kms$ so that 
$\langle \vth^2\rangle \ll \langle v_z^2\rangle \approx \sigma_z^2$.
For convenience, we shall use $v_z$ to denote the RMS value 
$\langle v_z^2 \rangle^{1/2}$.

The momentum flux difference also contains magnetic terms,
$\Delta P_{\rm mag}\equiv 
\Delta(B_x^2 + B_y^2 - B_z^2)/8\pi$,
where $\Delta$ denotes
the difference between midplane values and values
at the surface of the neutral-gas disk. Numerical simulations of
highly-supersonic magnetohydrodynamic turbulence
(e.g. \citealt{Stone98,MacLow98}) have shown that the magnetic
energy is typically smaller than the turbulent energy by a factor $\sim 2-3$, 
so that 
$B_{midplane}^2/(8\pi) \simlt \rho_0 \langle v_z^2\rangle $; this is also
consistent with observations \citep{Fer01}.  Since the magnetic scale
height $L_B$ is typically large compared to 
the semi-thickness $H\equiv \Sigma/\rho_0$ of the neutral-gas layer 
in galactic disks \citep{2000A&A...364L..36T,2009A&A...506.1123H}, 
the ratio $\Delta P_{\rm mag}/P_{\rm mag}\sim H/L_B\ll
1$, so that the resulting magnetic contribution to the momentum flux
difference will be small ($\simlt [H/L_B] \rho_0 \langle v_z^2\rangle$).
Cosmic rays and radiation may also contribute to the momentum flux difference,
with respective 
terms $\Delta P_{\rm cr}$ and $\Delta P_{\rm rad}$. 
Including all terms, the 
total momentum flux corresponds to an effective midplane
pressure $P_{eff}\equiv P_{\rm turb} + P_{\rm th} +
\Delta P_{\rm mag} +
\Delta P_{\rm cr} + 
\Delta P_{\rm rad}$.

By equating $P_{eff}$ with the total weight (the sum of equations
\ref{gas_weight} and \ref{bulgeweight}), we obtain 
\begin{eqnarray}
P_{eff}&\equiv& 
P_{\rm turb} + P_{\rm th} +
\Delta P_{\rm mag} +
\Delta P_{\rm cr} + 
\Delta P_{\rm rad} 
\equiv
\rho_0 \sigma_z^2  (1+ {\cal R})\nonumber \\
&=& \frac{\pi G \Sigma^2}{2}\left(1+ \chi \right).
\label{Pmid_eq}
\end{eqnarray}
Here, 
\begin{equation}
{\cal R} \equiv \frac{\Delta P_{\rm mag} +
\Delta P_{\rm cr} + 
\Delta P_{\rm rad}}{\rho_0 \sigma_z^2}
\label{Rdef}
\end{equation}
characterizes the contribution to vertical support from 
non-kinetic compared to kinetic terms.  The term 
$\chi \equiv  4 \zeta_d \rho_b /(3 \rho_0)$ represents
the contribution to the gas disk's 
weight from the external bulge potential compared to 
the gas disk's self-gravitational weight.
Since 
the midplane gas density is given by $\rho_0= 4 \zeta_d \rho_b/(3 \chi)$, 
the vertical equilibrium equation (\ref{Pmid_eq}) is a quadratic in $\chi$ 
that can be solved to obtain 
\begin{equation}
\chi = \frac{2C(1+{\cal R})}{1+ \left[1+4C(1+{\cal R})\right]^{1/2}}, 
\label{chi_def}
\end{equation}
where $C\equiv 8 \zeta_d \rho_b \sigma_z^2/(3\pi G \Sigma^2)$ depends only 
on large-scale properties of the system.  Note 
that $C^{1/2}$ is approximately equal to the ratio of scale heights for
a disk-only potential compared to a bulge-only potential, so that
$C \ll 1$  if the gas disk dominates vertical gravity 
and $C \gg 1$ if the bulge dominates vertical gravity.

Equation (\ref{chi_def}) can also be rearranged to yield 
$C=\chi(1+\chi)/(1+{\cal R)}$; 
when ${\cal R}, C\ll 1$, $\chi \approx C$.
Using $\rho_b= 3\Omega^2/(4\pi G)$, 
$C= 2\zeta_d{\cal W}^2 \approx 0.7 {\cal W}^2$ 
in terms of the quantity 
${\cal W}\equiv (\sigma_z \Omega )/(\pi G \Sigma)$.
The parameter $\cal W$ is 
the analog of the \citet{Toomre64} parameter $Q\equiv \kappa 
\sigma_R/(\pi G \Sigma)$
for vertical 
rather than horizontal oscillation frequency and velocity dispersion; i.e. 
${\cal W}= Q\sigma_z/(2\sigma_R)$ for epicyclic frequency $\kappa=2\Omega$. 
Thus, if $Q\sim 1-2$, ${\cal R}\ll 1$, and $\sigma_z/\sigma_R \simlt 1$,
then $C \simlt 0.2 -0.7$ and  
$\chi \simlt 0.2-0.5$ so that self-gravity dominates
the external (stellar bulge) gravity in controlling the vertical pressure, 
and $\rho_0\simgt \rho_b$.
Alternatively, 
\begin{equation}
Q=\frac{ \sigma_R }{\sigma_z  }\left(\frac{2  }{\zeta_d}  \right)^{1/2}
\left[\frac{\chi(1+\chi)}{1+{\cal R}  }  \right]^{1/2},
\label{Toomre_eq}
\end{equation}
so that for ${\cal R}\ll 1$ and $\chi \simlt 1$, 
$Q \simlt 3 \chi^{1/2}\sigma_R/\sigma_z$.  Also, note that the bulge potential
generally dominates orbital motion (even if the gas potential dominates 
vertical motion), with a factor $\sim (R/H) C$ for the stellar vs.
gas contributions to $\Omega^2$.

If star formation is self-regulated via momentum inputs, 
we expect the turbulent portion of the 
vertical momentum flux through the disk, 
$\rho_0 \langle v_z^2 \rangle$,
to be comparable to the total vertical momentum per unit time per
unit area that is injected in the gas by feedback from star formation.  For
isotropic momentum $p_*$  injected per massive star near the
midplane, averaging over spherical shells 
yields a vertical component injected on each side of the disk of 
$p_*/4$.  We shall therefore adopt the relation  
\begin{equation}
P_{\rm turb}=\rho_0 \langle v_z^2\rangle = f_p \frac{p_*}{4m_*}\sigsfr,
\label{turbflux}
\end{equation}
 where $m_*$ is the total mass in stars
formed per massive star.  The ratio $p_*/m_*$ is simply the mean radial
momentum injected in the ISM per unit mass of stars formed.  
We have introduced a factor $f_p$ to parameterize the dependence 
on the details
of turbulent injection and dissipation.  The value of $f_p$ is expected to vary
between 1 (for strong dissipation) and 2 (for weak dissipation; this
is the limit for a stream of particles that is injected vertically
and conserves energy as the particles fall back to the midplane). 

The kinetic momentum injected in the ISM per massive star can have
contributions from a number of sources, including expanding \ion{H}{2}
regions, stellar winds, and supernovae
(e.g. \citealt{ 1996ApJ...467..280N,2004RvMP...76..125M}).  If star
clusters are born in sufficiently optically-thick regions with deep
gravitational potential wells, radiation pressure also becomes important
in accelerating the residual gas to high
velocities \citep{2010ApJ...709..191M,2009ApJ...703.1352K}.
\citet{2002ApJ...566..302M} estimates the momentum injected by an
expanding \ion{H}{2} region for a given source of ionizing luminosity
and ambient density within a GMC, finding a ratio of momentum-to-mass
$p_*/m_*\sim 200 - 300 \kms$ for clouds of mass $\sim 10^5
-10^7\Msun$.  Using the results of 
\citet{2002ApJ...566..302M}, it can be shown that $p_*/m_*  \propto 
M_{\rm GMC}^{-4/7} R_{\rm GMC }^{3/7} \varepsilon_{\rm GMC}^{-3/7}$ for 
$M_{\rm GMC}$, $R_{\rm GMC }$, and 
$\varepsilon_{\rm GMC}$ the mass, radius, and 
integrated 
star-forming efficiency of a GMC, so that GMCs with 
escape speeds 
$(G M_{\rm GMC}/R_{\rm GMC })^{1/2}> 9 \kms$ or $\varepsilon_{\rm GMC}>0.01$ 
would have $p_*/m_*\simlt 200 \kms$.
Radiation pressure is most important when gas is concentrated near a
central stellar cluster and the efficiency is high 
(see Appendix A), so that the value of $p_*/m_*$
will be comparable to the escape speed from the (super)cluster that forms --
up to $\sim 100 \kms$ for the most extreme clusters 
\citep{2003ApJ...596..240M,2004A&A...416..467M,2009ApJ...706..203O}.

Although radiation and ionized-gas momentum inputs are important for destroying 
individual gravitationally-bound GMCs (thereby limiting star formation),
supernovae are likely the most important feedback mechanism 
for driving turbulence in the ISM as a whole \citep{1978ppim.book.....S}.
For supernova energy of $10^{51}\, {\rm  erg}$ 
and ambient density $10^3-1 \pcc$,
the momentum injection per event is 
$p_* \approx 2-5\times 10^5 \Msun \kms$ 
\citep{Chevalier74,Cioffietal88}.  If multiple supernovae combine to drive a 
single expanding shell, 
stellar winds contribute as well, but this increases the total 
energy (and momentum, which is approximately linear in the input energy) 
per unit mass by less than $\sim 20\%$ \citep{1999ApJS..123....3L}.
For a \citet{Kroupa01} IMF, the total mass in
stars per high mass star ($M>8\Msun$) is $m_*=100\Msun$, so we shall  
adopt $p_*/m_* = 3000 \kms$ from supernovae as our fiducial numerical value. 

Under the assumption that cosmic rays are accelerated by supernova blast waves
and diffuse vertically out of the disk,
the momentum flux contribution from cosmic rays will be proportional to 
that from supernovae, with 
\begin{equation}
\Delta P_{\rm cr}=\frac{H}{L_{\rm cr}} 
\frac{ E_{\rm SN} \zeta_{\rm cr}}{2 v_A m_* }\sigsfr.
\end{equation}
Here, 
$\zeta_{\rm cr}$ is the efficiency of cosmic ray production 
per supernova of energy $E_{\rm SN}$, and we have assumed that the cosmic 
ray fluid streams at speed $v_A$.
The mean free path of cosmic rays, $L_{\rm cr}$, is likely comparable 
the magnetic scale height $L_B$ \citep{2008ApJ...673..942Y}.  
Taking $\zeta_{\rm cr} \sim 0.1$ 
\citep{2008ARA&A..46...89R},  $v_A\sim 10\kms$, 
and $E_{\rm SN} =10^{51}\, {\rm erg}$, 
$(E_{\rm SN} \zeta_{\rm cr})/(2 v_A m_*) =2500\kms$; this is comparable to 
the kinetic input $p_*/m_*$ from supernovae.  
Since the neutral-disk thickness is small compared to the 
magnetic scale height, however, 
$H/L_{\rm cr} \sim H/L_B \ll 1$ and  
the vertical support of the neutral disk from cosmic rays is small compared to 
that from turbulence given in equation (\ref{turbflux}), 
$\Delta P_{\rm cr}/(\rho_0 \langle v_z^2\rangle )\sim H/L_{\rm cr}\ll 1$, 
similar to the situation for magnetic support 
$\Delta P_{\rm mag}/(\rho_0 \langle v_z^2\rangle )\sim H/L_{\rm B}\ll 1$.

Radiation pressure is important to vertical support of the disk if it
is optically thick to reprocessed infrared radiation \citep{TQM05}. Assuming
uniformly-distributed sources at the disk midplane, the
radiation pressure term is 
\begin{equation}
\Delta P_{\rm rad}= \frac{\kappa_{\rm IR}\Sigma }{2}\frac{F_{\rm rad}}{2 c } 
= \frac{\epsilon_* c \kappa_{\rm IR} \Sigma }{4  } \sigsfr . 
\label{radflux}
\end{equation}
Here,  $\kappa_{IR}$
is the mean opacity,
$F_{\rm rad}$ is the total luminosity per unit area produced by the 
disk (half is radiated from each side),
$c$ is the speed of light, 
and based on a standard Starburst99 model \citep{1999ApJS..123....3L},
$\epsilon_* \equiv L_*/(c^2 \dot M_*)\approx 6.2\times 10^{-4}$ 
is the mass-to-radiation energy conversion efficiency by stars for a 
\citet{Kroupa01} IMF in steady state  
(applicable if the starburst duration is $\simgt 10^7 \yr$).
The radiation pressure term 
becomes comparable to the kinetic term driven by supernovae 
(eq. \ref{turbflux}) if 
$\epsilon_* c\kappa_{\rm IR} \Sigma $ approaches $p_*/m_*$.

Retaining just the turbulence and radiation pressure terms 
in equation (\ref{Pmid_eq}), 
and using equations (\ref{turbflux}) and (\ref{radflux}),  the surface
density of star formation is given by
\begin{equation}
\sigsfr =  \frac{2\pi }{f_P }
\frac{\left(1+ \chi   \right)   }{1 + \tau/\tau_*}
\frac{m_* G   \Sigma^2}{p_*}
 .
\label{Sigma_SFR_P_eq}
\end{equation}
Here, 
\begin{equation}
\tau_*\equiv \frac{f_p p_*  }{\epsilon_* m_*  c} 
=16 f_p 
\left(\frac{p_*/m_*}{3000 \kms  }\right)
\left(\frac{\epsilon_*}{6.2\times 10^{-4}}\right)^{-1},
\label{tau*_eq}
\end{equation}
and the optical depth through the disk is 
\begin{equation}
\tau \equiv \kappa_{\rm IR} \Sigma
=0.21
\left(\frac{\kappa_{\rm IR}  }{10 \cm^2 \g^{-1}  }  \right)
\left(\frac{\Sigma }{100 \Msun \pc^{-2}  }  \right).
\label{tau_eq}
\end{equation}
Trapped radiation begins to affect the star formation rate 
when 
$\tau/\tau_* \simgt 1$, for 
very high gaseous surface densities.\footnote{
Streaming 
radiation, as well as diffusing radiation, injects 
momentum to the disk when it is absorbed.  The maximum momentum flux 
that can be injected to each side of the disk 
by streaming stellar radiation (including ionizing and far-UV
radiation before it is reprocessed to IR) is the input value, 
$F_{\rm rad}/(2c)$.  The maximum effect of streaming radiation on 
self-regulating disk star formation is obtained by replacing 
$\kappa_{\rm IR}\Sigma /2=\tau/2$ by 1 in equation (\ref{radflux}), so that 
$\tau/\tau_*\rightarrow 2/\tau_*=2\epsilon_* m_*  c  /(f_p p_*)$ 
in equation (\ref{Sigma_SFR_P_eq}).  Using the
fiducial value of $\tau_*$ given in equation (\ref{tau*_eq}), 
the maximum effect of momentum input from streaming stellar radiation is 
$\sim 10\%$ compared to that from supernovae.
}
Note that  in equation (\ref{Rdef}), ${\cal R} \rightarrow \tau/\tau_*$ 
if radiation dominates over the magnetic 
and cosmic-ray terms and turbulence dominates the thermal term.

In the turbulence-dominated regime ($\tau/\tau_* \ll 1$) for self-regulated 
disks, the surface density of star formation is 
\begin{eqnarray}
\Sigma_{\rm SFR,turb} &=& 
  \frac{2\pi \left(1+ \chi   \right)}{f_P }
\frac{ m_* G   \Sigma^2}{p_*  }
\nonumber \\
\nobreak
&=&0.092 \Msun \kpc^{-2} \yr^{-1}
\frac{\left(1+ \chi   \right)}{f_p} 
\left(\frac{p_*/m_* }{3000 \kms} \right)^{-1}
\left(\frac{\Sigma }{100\Msun\pc^{-2} } \right)^2.
\label{Sigma_SFR_turb}
\end{eqnarray}
As noted above, disks that are marginally gravitationally unstable
($Q\simlt 2$) have $\chi \simlt 0.5$.  Thus, 
turbulence-controlled, 
self-regulated star formation in galactic centers is expected to follow a 
scaling relation $\sigsfr \propto \Sigma^2$.

At very high surface densities such that $\tau/\tau_* \gg 1$ (and
assuming $\chi \ll 1$), the star formation rate in the
radiation-pressure-dominated regime is
\begin{eqnarray}
\Sigma_{\rm SFR, rad} &=& \frac{2\pi G \Sigma  }{\epsilon_* c\kappa_{\rm IR}  }
\nonumber\\
&=& 710 \Msun \kpc^{-2} \yr^{-1} 
\left(\frac{\epsilon_*}{6.2\times 10^{-4}}\right)^{-1}
\left(\frac{\kappa_{\rm IR}}{10 \cm^2 \g^{-1}}\right)^{-1}
\left(\frac{\Sigma  }{10^4 \Msun\pc^{-2}  }\right),\hskip .2 in 
\label{Sigma_SFR_rad}
\end{eqnarray}
which is linear rather than quadratic in the gas surface 
density.

Equation (\ref{Sigma_SFR_turb}) would apply to galactic center regions
experiencing all but the most extreme starburst activity, with a transition 
to the radiation-dominated regime of equation (\ref{Sigma_SFR_rad}) at 
$\tau \sim \tau_*$, corresponding to a gas surface density 
\begin{equation}
\Sigma_{\rm trans, rad}=  8\times 10^3 \Msun \pc^{-2} 
\left(\frac{\kappa_{IR}}{
10 \cm^2 \g^{-1}}\right)^{-1}
\left(\frac{p_*/m_*}{3000 \kms  }\right)
\left(\frac{\epsilon_*}{6.2\times 10^{-4}}\right)^{-1}.
\end{equation}
The self-regulated turbulent galactic-center regime of equation 
(\ref{Sigma_SFR_turb}) 
connects with the ``mid-disk'' regime, where internal GMC processes control
star formation (see Section 1), at lower surface densities.  Setting 
equation (\ref{Sigma_SFR_turb}) equal to 
$\sigsfr = \Sigma/t_{\rm SF, GMC}$ with $t_{\rm SF, GMC}$ 
comparable to the empirical mid-disk value
$t_{\rm SF, mol} \sim 2 \Gyr$ \citep{Bigiel08,2009ApJ...704..842B}
and assuming $\chi \ll 1$, the nominal 
transition to the mid-disk regime is at 
\begin{equation}
\Sigma_{\rm trans, GMC}
= 54 \Msun \pc^{-2} f_p  \left(\frac{p_*/m_*}{3000 \kms  }\right)
\left(\frac{t_{\rm SF,GMC}}{2 \Gyr }\right)^{-1}.
\end{equation}
Physically, this transition occurs because at low surface densities, 
the energy input from star formation is insufficient to prevent 
``diffuse'' 
molecular gas from becoming concentrated in individual gravitationally-bound 
GMCs.


\section{Comparison of $\sigsfr$ vs. $\Sigma$ to numerical simulations}

To initiate a numerical study of star formation under high surface
density conditions, we begin with a very simple computational model,
which nevertheless includes sufficient ingredients that we are able to
investigate self-regulation of star formation in turbulence-dominated
disks.  Our simulation domain represents a local region within a
starburst disk, which we resolve both vertically and horizontally.
The domain is rotating with the disk, so we include centrifugal and
Coriolis forces; we assume a solid-body rotation curve, so there is no
large-scale rotational shear within the domain.  We assume that the
gas cools efficiently to maintain an approximately constant (low)
temperature, so we adopt an isothermal equation of state with sound
speed $\cs=(P/\rho)^{1/2}$.  For this first study, we neglect magnetic
fields.  Thus, the dynamical equations we solve are:
\begin{equation}\label{eq:cont}
 \frac{\partial\rho}{\partial t}+\nabla\cdot(\rho \vel)=0,
\end{equation}
\begin{equation}\label{eq:mom}
 \frac{\partial\vel}{\partial t}+\vel\cdot\nabla\vel=
 -\frac{1}{\rho}\nabla P - 2\mathbf{\Omega}\times
 \vel-\nabla\Phi_g +\mathbf{g}_{\rm ext},
\end{equation}
and
\begin{equation}\label{eq:poisson}
\nabla^2\Phi_g=4\pi G\rho
\end{equation}
(e.g. \citealt{2002ApJ...581.1080K}, taking dimensionless shear
parameter $q=0$).  For a spherical bulge, the vertical gravitational 
field is given by $\mathbf{g}_{\rm ext} = - \Omega^2 z \hat z$.  

We
integrate the dynamical equations using a version of the {\it Athena}
code \citep{2008ApJS..178..137S}, which employs a single-step,
directionally-unsplit Godunov scheme
\citep{2005JCoPh.205..509G,2008JCoPh.227.4123G} to obtain
conservative, second-order accurate solutions. The Poisson equation is
solved using the Fourier transform method of \citet{KO09a}, suitable for disk
problems that are treated as having open boundary conditions in the
vertical ($\hat z$) direction and periodic boundary conditions in the
horizontal directions.  The equations above can be solved either 
for a three-dimensional domain, or a two-dimensional domain 
representing a radial-vertical slice through the disk 
(e.g. \citealt{2006ApJ...649L..13K,KO09a}).
Two-dimensional radial-vertical models,
which we adopt here, capture the physics of the competition between 
turbulent driving and gravitational settling in the vertical direction,
and allow much more extensive initial exploration of parameters than is possible
for three-dimensional models (which we intend to pursue next).

Motivated by observations showing a linear relation between very dense gas and 
star formation 
(e.g. \citealt{2004ApJ...606..271G,2005ApJ...635L.173W,2010ApJS..188..313W}),
we assume that star formation takes place in gas above a density threshold 
$\rho_{\rm thr}=\mu \nthr $ at a fixed efficiency per free-fall time
$\epsff(\nthr)$; we also assume 
that the total mass in stars formed per massive star is
$m_*$, and that each massive star injects a momentum $p_*$ into the ISM.
Thus, when $n\ge \nthr$ in a given zone containing total mass $m$,
the probability of a massive star formation event occurring within a
given time step $\Delta t$ is set to be $P=\Delta t \epsff(\nthr)
m/[m_*t_{ff}(\rho)]$.  Here, $t_{\rm ff}(\rho)=[3 \pi/(32 G \rho)]^{1/2}$  
is the gravitational free-fall time at density $\rho$.
For every zone where a massive star formation event
occurs, we inject a total momentum $p_*$ into the gas in a
spherical region of radius $R_{\rm sh}$ surrounding the event.  For
every feedback event associated with a massive star, the tally of the
total mass in stars formed is increased by $m_*$, although we do not
remove any gas from the grid (in order to maintain constant $\Sigma$). 
The computed surface star formation rate
$\sigsfr$ is the mass in stars formed per unit time within the domain,
divided by the horizontal area.

A description of the detailed numerical methods and the results of a
full parameter survey (varying $\Sigma$, $\Omega$, $\cs$, $\epsff(\nthr)$,
$p_*/m_*$, $R_{\rm sh}$, $\nthr$, domain size $L_x$,
$L_z$, and numerical resolution), 
will be presented separately (Shetty \& Ostriker 2010, in
preparation).  Here, since we are interested in comparing to equation
(\ref{Sigma_SFR_turb}) to test the idea of turbulent self-regulation, we focus
on the dependence of the star formation rate $\sigsfr$ on the gas surface
density $\Sigma$.  We fix $m_*=100\Msun$, $p_*=3 \times 10^5 \Msun
\kms$, $R_{\rm sh} =5 \pc$, $\cs=2 \kms$, and set the angular velocity
to $\Omega = \Myr^{-1} (\Sigma/10^3\Msun \pc^{-2})$ (so that the
Toomre parameter would be unity for a velocity dispersion of $7
\kms$).  Note that the value of $\cs$ we use is larger than would be
provided by thermal pressure at temperatures $\sim 10-100\K$; without
magnetic fields (which are neglected in this study), shocks would
become unphysically strong if $\cs$ is much lower.  Turbulent energy
is still far larger than thermal energy for the parameter range under
consideration, so that the disk thickness is controlled by
feedback-driven turbulence (see Section 5 and Shetty \& Ostriker 2010, in
preparation). The models are integrated for $200 \Myr$; a
quasi-steady state typically requires only a few tens of $\Myr$ to be
established, for the parameter regime $\Sigma \simgt 100 \Msun
\pc^{-2}$ under investigation.

\begin{figure}[t]
\leftline{\psfig{figure=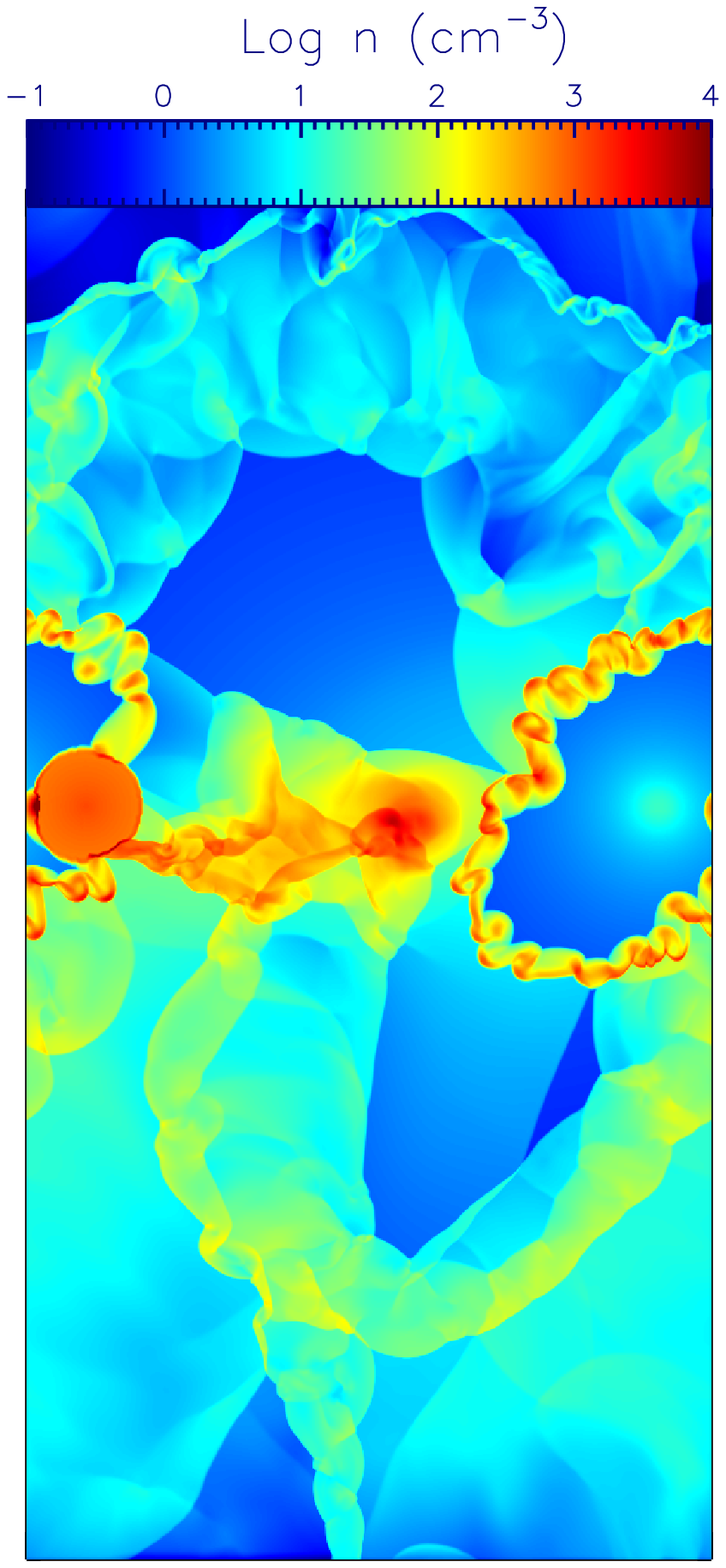,height=3truein}}
\vskip -3truein
\hskip 1.5truein
{\psfig{figure=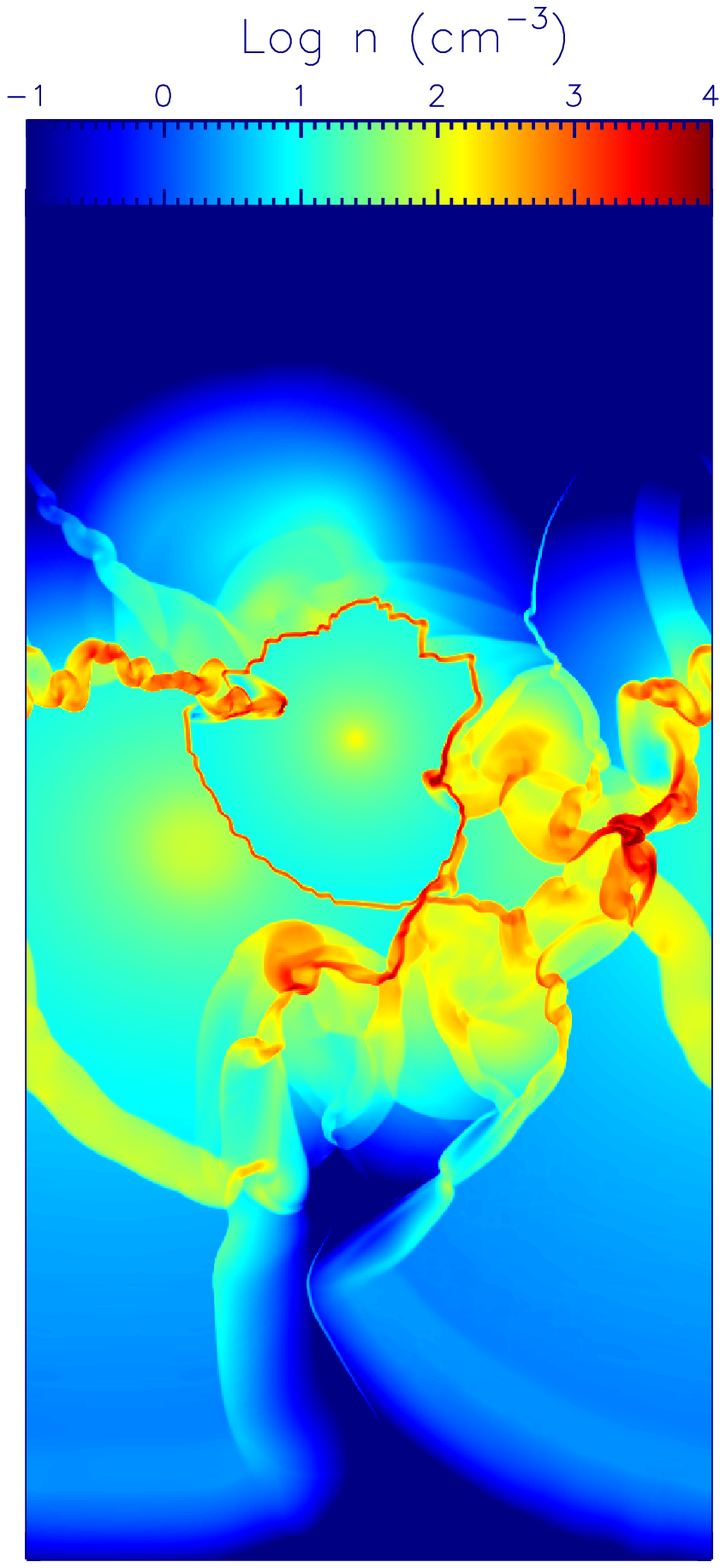,height=3truein}}
\vskip -3truein
\hskip 3.truein
{\psfig{figure=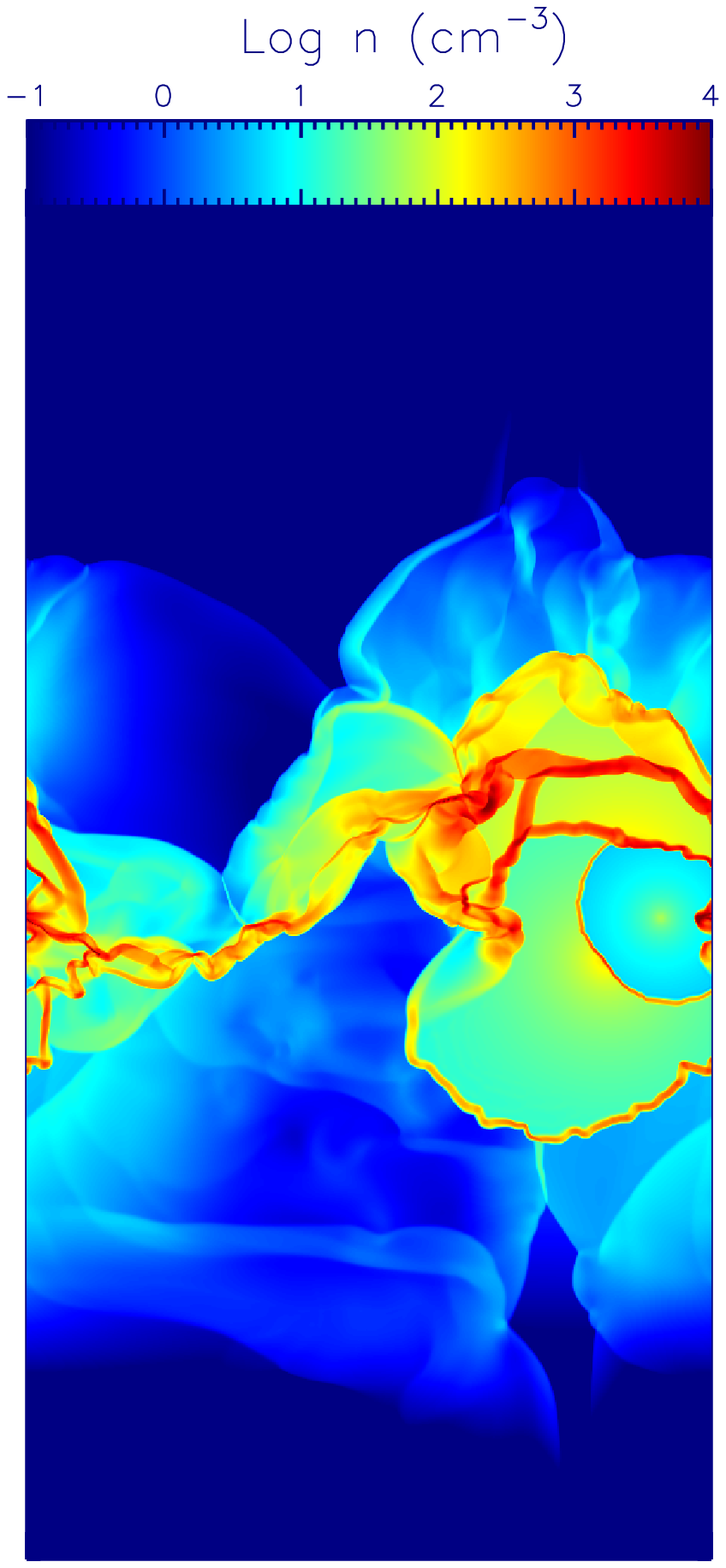,height=3truein   }}
\vskip -3truein
\hskip 4.5truein
{\psfig{figure=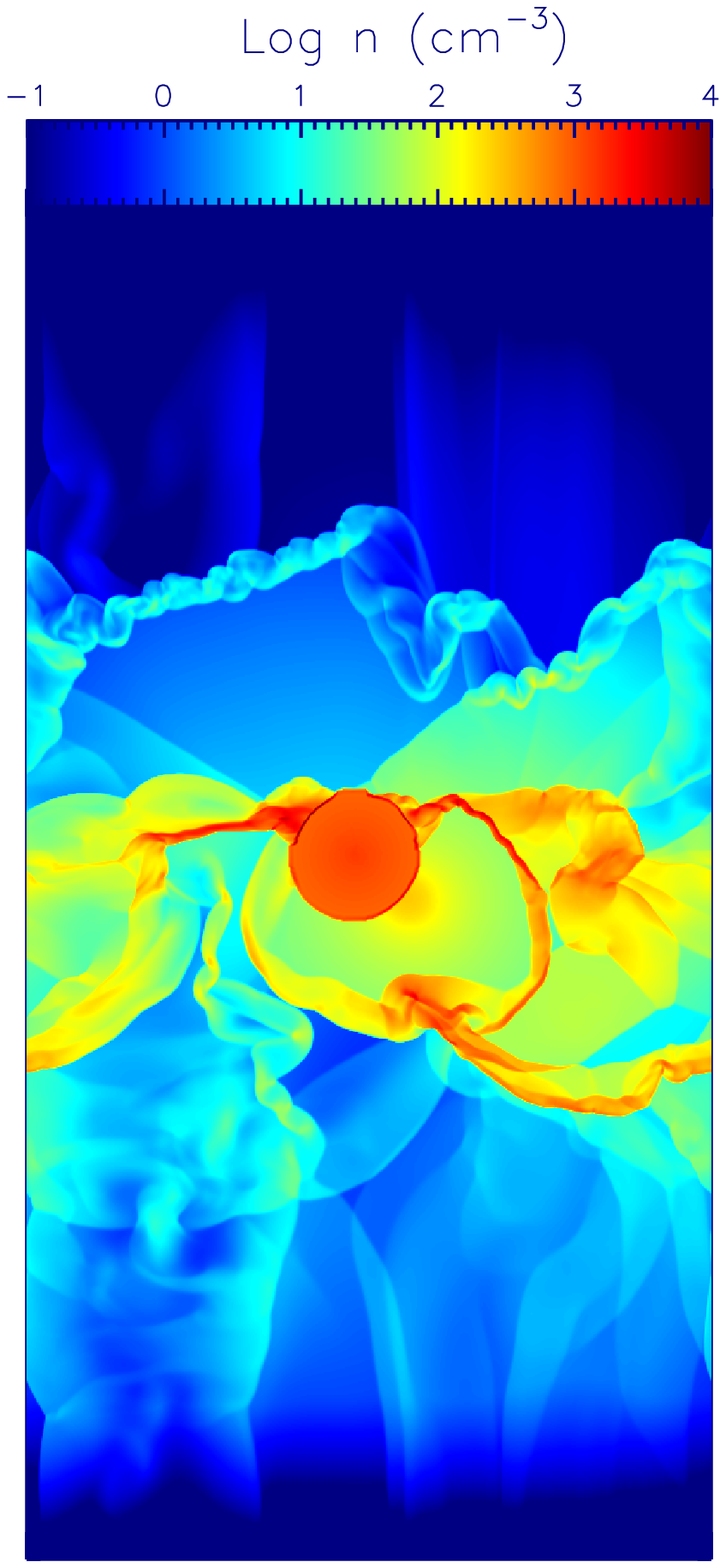,height=3truein   }}
\caption{Sample evolution of density structure (logarithmic color scale) 
in the radial-vertical plane, for 
a numerical model with $\Sigma=200 \Msun \pc^{-2}$. 
Snapshots are separated by $5 \Myr$.  The domain size is 
$L_x\times L_z = 60\times 120 \pc^2$.
   \label{fig2}}
\end{figure}

Figure (\ref{fig2}) shows an example of typical density structure
snapshots in the radial-vertical plane, from a model with $\Sigma =
200 \Msun \pc^{-2}$, taking $\nthr=5\times 10^3 \pcc$ and
$\epsff(\nthr)=0.01$.  This density threshold and efficiency factor are
chosen based on observations of Milky Way and extragalactic gas and
star formation (\citealt{KT07,2009ApJS..181..321E,2010arXiv1009.2985L,
  2010arXiv1009.1621H}; see also Section 5).  As Figure (\ref{fig2}) shows, the
structure is highly inhomogeneous due to the dominance of turbulence,
which shocks and compresses the gas.  Effects of recent star formation
events are seen as dense circular regions; these drive expanding shells, 
which are also evident.
Although structure is irregular and highly dynamic, gas is
preferentially concentrated toward the midplane due to gravity; this
is where star formation events take place, in the densest gas.  All of
our models show similar structure and evolution, with repeated local
collapse events and feedback-driven turbulent excitation.  In some of
our simulations with a sufficiently large radial domain, we find
that the gas may collapse into a single condensation that the feedback
is unable to redisperse, because the gravitational potential well is too
deep.  As we discuss in Section 5 and the Appendix, in real
galaxies radiation pressure may play a role in dispersing strongly
bound clouds and superclouds.  Since radiation is not included in the current
simulations, we consider only models that do not suffer this kind of global
collapse.

\begin{figure}[!ht]
\epsscale{.6}
\plotone{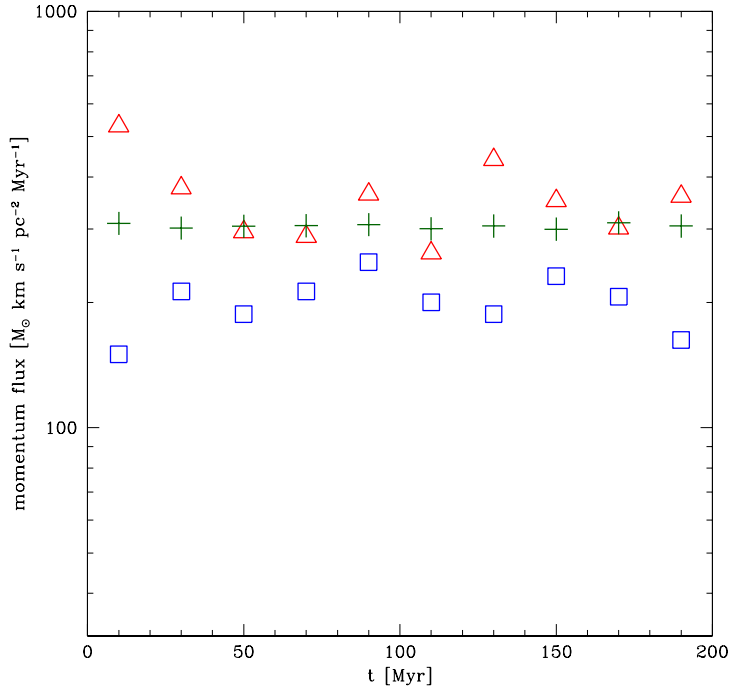}
\caption{Comparison of characteristic input vertical momentum flux from
 star formation $0.25(p_*/m_*)\Sigma_{\rm SFR}$  (squares), turbulent
 pressure in the vertical direction 
$\rho_0 v_z^2$ (triangles), and vertical weight of the gas disk 
$0.5\pi G \Sigma^2(1+\chi)$ (plusses),
for the same model as shown in Fig. (\ref{fig2}).
Each quantity is computed averaging over temporal bins of 20 Myr.
   \label{fig3}}
\end{figure}

\begin{figure}[!ht]
\epsscale{.60}
\plotone{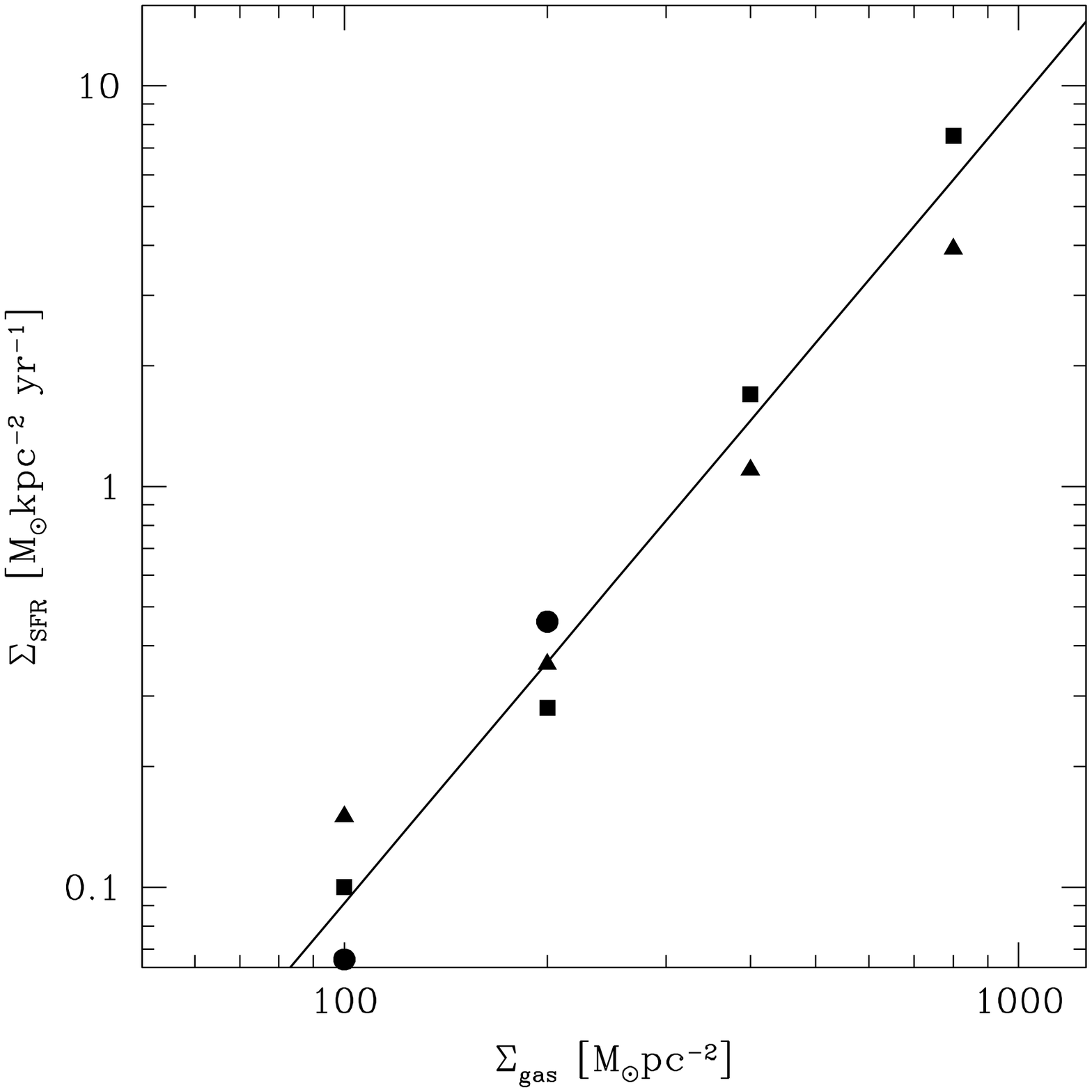}
\caption{Results of mean measured star formation rate $\sigsfr$ 
from a set of simulations with a range of gas surface density $\Sigma$, 
in comparison with the prediction given in 
equation (\ref{Sigma_SFR_turb}) (solid line).  For the
numerical results, triangles have $\epsilon_{\rm ff}(n_{\rm thr})=0.005$, 
squares have $\epsilon_{\rm ff}(n_{\rm thr})=0.01$, and 
circles have $\epsilon_{\rm ff}(n_{\rm thr})=0.05$.
\label{fig4}}
\end{figure}

In spite of their highly dynamic behavior, our simulations show that a
quasi-steady state is established in which the vertical weight of the
gas is approximately balanced by turbulent pressure, which itself is
driven by momentum injection associated with massive star formation.
For the same model shown in Figure (\ref{fig2}), Figure (\ref{fig3})
shows the evolution of the turbulent momentum flux in the vertical
direction $\rho_0 v_z^2$ and the characteristic vertical momentum
injection rate to each side of the disk from star formation $0.25
(p_*/m_*) \Sigma_{\rm SFR}$, as well as the mean vertical weight of
the gas disk $0.5 \pi G \Sigma^2(1+\chi)$ (the value of $\chi \sim 0.1$
for this model); the respective time-averaged values of these
quantities are 357, 209, and 305 $\Msun \kms \pc^{-2} \Myr^{-1}$.
From equation (\ref{turbflux}), $f_p \sim 1.7$, consistent with the
expectation that the turbulent pressure is comparable to the momentum
injection rate per unit area from star formation.  Our other models show similar
temporal behavior for the various momentum flux terms.

Figure (\ref{fig4}) shows the time-averaged values of $\sigsfr$
measured from a set of simulations with $\Sigma = 100, 200, 400, 800
\Msun \pc^{-2}$.  For all models we set $\nthr=5\times 10^3 \pcc$; we
show cases with $\epsff(\nthr)=0.005$, $0.01$, and $0.05$.  The numerical
results are consistent with the prediction given in equation
(\ref{Sigma_SFR_turb}) for momentum-controlled, self-regulated disk star
formation.  We note that both turbulence and small-scale self-gravity
are crucial in determining the star formation rate in these
simulations.  For all of the models, our adopted $\nthr$ exceeds the
midplane value of the density for a purely thermally supported slab
(with the adopted value $\cs=2 \kms$), so no star formation would take
place without localized gravitational collapse.  On the other hand, if
we had not imposed a high threshold density and had simply {\it
  assumed} a star formation rate equal to $\epsff \Sigma/\tff(n_{\rm
  midplane})$ for our adopted $\cs$ and $\epsff=0.01$ (see Section 5), 
we would have obtained the same scaling with surface density 
$\sigsfr \propto \Sigma^2$ as found with our simulations (and as 
predicted by equation [\ref{Sigma_SFR_turb}]), but a much
larger coefficient: $\sigsfr = \epsff 4 G \Sigma^2/(\sqrt{3}\cs)$.
For the value $\cs=2\kms$ used in the present simulations, this would yield
a factor of $\sim 6$ larger $\sigsfr$ than equation (\ref{Sigma_SFR_turb}); 
lower $\cs$ would yield even larger $\sigsfr$.
Thus, a simulation that did not include turbulent feedback but allowed
gas to cool to low temperature (small $\cs$) could significantly 
overestimate the star formation rate. Turbulence driven by star
formation keeps the volume-averaged midplane value of the density
lower than that of a thermally-supported disk, but self-gravitating
collapse of shocked, highly overdense local regions still 
occurs, leading to further (self-regulated) star formation.

\section{Comparison of $\sigsfr$ vs. $\Sigma$ to observations}

\begin{figure}[!ht]
\epsscale{.90}
\plotone{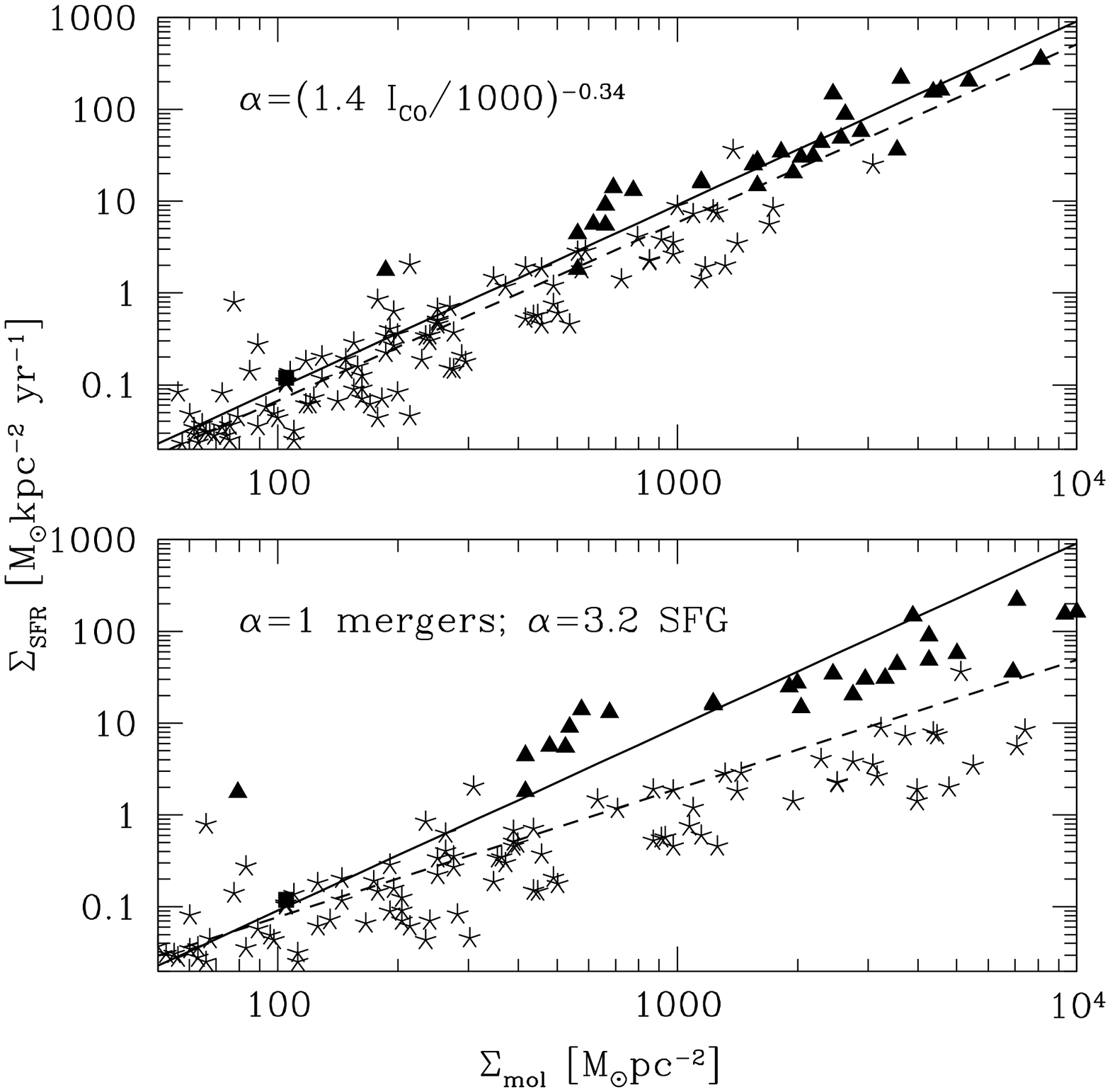}
\caption{Solution given by equation 
(\ref{Sigma_SFR_turb}) for star formation rate $\sigsfr$ 
vs. gas surface density $\Sigma$ ({solid line}), 
in comparison with data compiled in
\citet{2010MNRAS.407.2091G}.  
In the lower panel, 
$\Sigma_{\rm mol}/[\Msun \pc^{-2}] 
\equiv 1.4 \alpha I_{\rm CO}/[\K \kms \pc^{-2}]$ 
is computed assuming $\alpha=3.2$ for star-forming galaxies ({stars}) and 
$\alpha=1$ for merger systems ({triangles}), following 
\citet{2010MNRAS.407.2091G}.
In the upper panel,
a variable value of $\alpha$ is adopted such that $\alpha=3.2$ for 
$\Sigma=100 \Msun \pc^{-2}$ and $\alpha=1$ for $\Sigma= 1000 \Msun \pc^{-2}$.
Dashed lines show the least-squares fit to the data, with slopes 1.4 in the 
lower panel and 1.9 in the upper panel.
The value obtained by \citet{2009ApJ...702..178Y}
 for the Milky Way center  is also shown (squares in both panels).
    \label{fig5}}
\end{figure}

The dependence of $\sigsfr$ on $\Sigma$ predicted by the simple model
of Section 2 can be compared to observations.
\citet{2010MNRAS.407.2091G} have recently investigated the
relationship between molecular gas and star formation in a large
sample compiled from both local-Universe and high-redshift galaxies.
The compilation includes local ULIRGs as well as normal and merging
galaxies at $z\sim 1-3$.  Gaseous surface densities in this (and
other) data sets are based on observations of CO emission, so it is
necessary to convert the integrated line emission, $I_{\rm CO}$, to a
value of $\Sigma$. This conversion is accomplished using a factor
$\alpha \equiv \Sigma_{\rm H2}/I_{\rm CO}$ or $X_{\rm CO}\equiv N_{\rm
  H2}/I_{\rm CO}$, where $\Sigma_{\rm H2}=\Sigma/1.4$ is the surface
density of molecular hydrogen in $\Msun \pc^{-2}$ (the factor 1.4
accounts for helium in the total surface density), $N_{\rm H2}$ is the
molecular hydrogen column density in $\cm^{-2}$, and $I_{\rm CO}$ is
in units of $\K \kms$ so that $X_{\rm CO} = 6.3 \times 10^{19} \alpha$
in units $\cm^{-2}/( \K \kms)$ if $\alpha$ has units $\Msun
\pc^{-2}/(\K \kms)$.  In Milky Way and Local Group GMCs, values of
$\alpha\approx 2-5$ (for the CO J=1-0 transition) have been inferred
based on a variety of empirical techniques
(e.g. \citealt{2007prpl.conf...81B}), whereas there is evidence that
$\alpha$ is significantly lower in ULIRGs and other extreme
star-forming systems (e.g. \citealt{2005ARA&A..43..677S}).

For surface
densities $\Sigma\simlt 100 \Msun \pc^{-2}$ in galaxies, CO emission can be
understood as arising primarily from collections of gravitationally bound 
GMCs
with limited variation in individual properties such as total column
density and kinetic temperature \citep{1987ApJ...319..730S,Bol08,
  2009ApJ...699.1092H}, such that there is an approximately constant conversion
factor $\alpha$. The standard value of $X_{\rm CO}= 2 \times 10^{20}$ 
based on CO emission in the Milky Way
\citep{1996A&A...308L..21S,2001ApJ...547..792D} corresponds to
$\alpha=3.2$.  Shetty et al (2010) have applied radiative transfer to 
numerical simulations of turbulent clouds, confirming that mean ratio 
of $N_{\rm H2}/I_{\rm CO}$ is comparable to standard empirically-estimated 
values $X_{\rm CO}= 2 \times 10^{20} $, provided that $A_V$ 
is neither very low nor very high.

For galactic center regions with high surface densities, however, the
molecular gas may respond to the overall gravitational potential in
the disk rather than being bound in individual GMCs (raising the velocity 
dispersion), and may be
systematically warmer for higher $\sigsfr$.  These effects would
increase the integrated CO emission $I_{\rm CO}$ for a given $\Sigma$, 
implying a smaller $X_{\rm CO}$
or $\alpha$ is needed \citep{1997ApJ...478..144S}.  
Since the dependence of $\sigsfr$ on
$\Sigma$ is nonlinear, gas surface densities may also be overestimated
in some cases if the gas is not spatially resolved and the
characteristic radius of the IR-emitting starburst is adopted as a
surrogate (this spatial mismatch is known to be an issue for
high-redshift galaxies; see e.g. \citealt{2009arXiv0912.1598B}).  In 
addition, lack of spatial resolution combined with galactic inclination 
can broaden observed CO line widths (due to rotation and to radial and/or 
azimuthal streaming within a beam) 
beyond the true turbulent velocity width in a local patch of the disk; 
without detailed modeling, it is not clear how this would affect the 
conversion between $I_{\rm CO}$ and $N_{\rm H2}$.
For ULIRGs with typical $\Sigma \sim 10^3\Msun \pc^{-2}$, empirically-estimated 
values are $\alpha \simlt 1$ \citep{1998ApJ...507..615D}.

From a theoretical point of view, the value of $\alpha$ in a
self-regulated starburst system should depend primarily on the disk's gas
surface density $\Sigma$, since this sets $\sigsfr$ and thus the gas
velocity dispersion and the distributions of
temperature and density through star formation feedback.  Using
numerical simulations for a range of disk properties combined with
detailed radiative transfer calculations, it will ultimately be
possible to obtain predictions for the dependence of $\alpha$ on local
conditions.  Empirically, a systematic trend for $\alpha$ to decrease
with increasing $\Sigma$ is already evident
(e.g. \citealt{2008ApJ...680..246T}).  The simplest possible relation
would be for $I_{\rm CO}$, and therefore $\alpha$, to depend on
$\Sigma$ as a power law.  For example, if $\alpha$ is fixed to 1 at
$\Sigma=10^3 \Msun\pc^2$ from ULIRG observations, and to 3.2 at
$\Sigma=10^2 \Msun\pc^2$ from the limit of 
``normal'' GMCs, a
power-law assumption yields $\alpha =(\Sigma/1000)^{-0.52}= (1.4
I_{\rm CO}/1000)^{-0.34}$.

In Figure (\ref{fig5}), we compare the prediction of equation
(\ref{Sigma_SFR_turb}) with the observed relation between $\Sigma$ and
$\sigsfr$  from the compilation of \citet{2010MNRAS.407.2091G}.
Also shown is the surface density and star formation rate in the Milky Way
Galactic center, from \citet{2009ApJ...702..178Y}.
For the theoretical result,
we have set $\chi \rightarrow 0$ in equation (\ref{Sigma_SFR_turb}) from  
equation (\ref{chi_def}), since  
$C<0.1$ in the observed sample 
(except for two cases which have $C <0.5$) if we assume the vertical velocity 
dispersion is $\sim 10 \kms$.
In the presentation of \citet{2010MNRAS.407.2091G}, a value
$\alpha=1$ is adopted for merger galaxies (LIRGs and ULIRGs), while a
value $\alpha=3.2$ is adopted for other systems.  In Figure 
(\ref{fig5}), we compare to observations using both these standard conversion
factors ({\it lower panel}), and using the simple power-law function for 
$\alpha$ given above ({\it upper panel}).

Evidently, the theoretical
prediction of equation
(\ref{Sigma_SFR_turb}) follows the data 
over the full range of 
surface densities if a $\Sigma$-dependent $\alpha$ is used.
If $\alpha=1$ is used for mergers and $\alpha=3.2$ for normal galaxies, 
equation (\ref{Sigma_SFR_turb}) agrees with the data at low $\Sigma$, but 
at high $\Sigma$ there is an offset between the two classes of galaxies,
with only the mergers close to the result of equation (\ref{Sigma_SFR_turb}). 
\citet{2010MNRAS.407.2091G} and 
\citet{2010ApJ...714L.118D} previously noted this offset, and suggested 
that a difference in dynamical timescale $\Omega^{-1}$ may be responsible 
for the apparent difference in star formation laws. With 
$\alpha$ depending on $\Sigma$, star formation follows a single relation 
for both mergers and normal galaxies in the $\Sigma-\sigsfr$ plane, 
but mergers are preferentially at 
higher $\Sigma$ -- which is correlated empirically with higher $\Omega$.
Regardless of the value of $\alpha$, star formation is observed to be 
more efficient in merger systems, based on the ratio 
$L_{\rm IR}/L_{\rm CO}$. Given the characteristically larger values of 
$I_{\rm CO}$ and $\Sigma$ for mergers compared to normal galaxies, 
this is consistent with theoretical expectations that $\sigsfr$ depends 
nonlinearly on $\Sigma$ -- in particular, as $\sigsfr\propto\Sigma^2$ 
based on equation (\ref{Sigma_SFR_turb}).

\section{Disk properties in the turbulence-dominated regime}

In Section 2, we showed that if turbulence driven by star formation
feedback dominates other forms of pressure, the equilibrium star
formation rate is expected to be proportional to the weight of the
ISM.\footnote{The star formation rate is also expected to be
  proportional to the weight of the ISM if thermal pressure dominates
  other forms of pressure, in a medium dominated by diffuse atomic gas
  and heated by young stars \citep{Ostriker2010}.  If, however,
  radiation pressure dominates both turbulent and thermal pressure,
  the star formation rate would be proportional to the vertical
  gravitational {\it field} rather than the {\it weight} (i.e. reduced
  by a factor $\propto 1/\tau\propto 1/\Sigma$ -- cf.  equations
  [\ref{Sigma_SFR_turb}] and [\ref{Sigma_SFR_rad}]).}  The force
balance (or momentum conservation) argument leads to a relation
$\sigsfr \propto \Sigma^2$ for gas-dominated regions in a
self-regulated steady state, but it says nothing about the internal
processes or character of the disk.  In order to maintain equilibrium,
the disk must internally adjust itself so that stars are produced at
the required rate, and this internal arrangement must also be
consistent with the detailed dynamics of a self-gravitating, turbulent
hydrodynamic (or hydromagnetic) flow.

In a region of mean gas 
density $\bar \rho$ (and where $\rho_*\ll \bar\rho$), 
the gravitational free-fall time at the
mean density, $t_{\rm ff}(\bar\rho)=[3 \pi/(32 G \bar\rho)]^{1/2}$,
represents a characteristic dynamical time.  The star formation rate
can then be expressed as $\dot M_* \equiv \epsff(\bar \rho) M_{\rm
  gas}/ \tff(\bar\rho)$, where $\epsff(\bar \rho)$ is an efficiency
factor that depends on the detailed properties of the gas.  
  Although it is uncertain exactly what controls the
rate of localized gas collapse in highly-turbulent systems, evidence
from current observations \citep{KT07,2009ApJS..181..321E}, theory
\citep{KM05,2009arXiv0907.0248P}, and numerical simulations
\citep{2009arXiv0907.0248P,2009arXiv0911.1795R,2009ApJ...707.1023V}
suggests that the star formation efficiency factor is low,
$\epsff(\bar \rho) \sim 0.01$, for molecular gas in observed
star-forming regions at a range of $\bar\rho$.  \citet{KM05} proposed
that for turbulence-dominated systems, $\epsff(\bar \rho)$ is given by
the fraction of mass compressed to densities higher than the mean by a
factor $\sim{\cal M}^2$ for ${\cal M}\equiv v_{\rm turb}/\vth$ the
Mach number. Because the density PDF in
turbulent regions is a lognormal function of $\cal M$
\citep{2007ARA&A..45..565M}, \citet{KM05} show that the resulting
dependence of the star-forming efficiency 
on the Mach number is weak,
$\epsff(\bar \rho) \propto {\cal M}^{-0.3}$ .  Thus, if localized 
collapse to make
stars is controlled primarily by turbulence, the overall rate is
expected to be 
governed primarily by the free-fall time on large scales.

In (self-gravitating) disks, it is convenient to use the mean midplane
density $\rho_0=\mu n_0$ as a reference value, so that the star
formation rate per unit area can be written as
\begin{equation}
\sigsfr\equiv \epsff(\rho_0)\frac{\Sigma}{ \tff(\rho_0)}.
\label{SFR-tff-eq}
\end{equation}
For a Gaussian vertical density distribution, $\bar \rho=\rho_0/\sqrt{2}$ 
so that $\epsff(\rho_0)=0.8 \epsff(\bar\rho)$.
As in Section 2, in adopting the form (\ref{SFR-tff-eq}), we have
implicitly assumed that the gas collapses only locally -- on scales
small compared to the Toomre wavelength \citep{Toomre64}
$\lambda_T\equiv G \Sigma (2 \pi/\kappa)^2$ -- so that angular
momentum and shear do not dominate the dynamics of star formation.  In
addition, equation (\ref{SFR-tff-eq}) implicitly assumes that the
free-fall time at the mean midplane gas density is shorter than the
vertical oscillation time in the gravitational field of the stellar
bulge; this will be true if gas self-gravity dominates vertical potential
gradients.

We note that equation (\ref{SFR-tff-eq}) is {\it not} expected to
apply in regions of a galactic disk that have a fundamentally
multiphase character, in terms of the mass- and volume- fractions of
gas at different temperatures. In particular, if most of the ISM's
mass is in clouds that are self-gravitating entities isolated from
their surroundings, the local density within these self-gravitating,
star-forming clouds need not be proportional to the (volume-weighted)
mean midplane density of the disk.  
Using multiphase disk simulations, \citet{KO09a}
indeed found that equation (\ref{SFR-tff-eq}) would yield a different
(steeper) star formation law than is obtained by adopting a constant
star formation timescale only in very dense (gravitationally-bound)
gas. 

For outer-disk regions dominated in mass by diffuse atomic gas,
\citet{Ostriker2010} argued that a key factor regulating star
formation is the heating and cooling of the warm, volume-filling
medium.  Gas that cannot be maintained in thermal equilibrium in the
diffuse phase cools and drops out to make gravitationally-bound, star-forming
clouds. This increases the heating rate (from stellar UV), and 
decreases the cooling rate (since the density is lower), 
for the remaining diffuse gas.  Mass dropout continues until
heating balances cooling.  In a state of simultaneous thermal and
dynamical equilibrium, the star formation rate for an outer-disk
region will still vary inversely with the dynamical time of the
dominant gas component, analogous to equation (\ref{SFR-tff-eq}), but
the relevant dynamical time is not $\tff(\rho_0)$. In outer-disk
regions, the ISM scale height is generally set by the gravity of stars
plus dark matter, rather than the self-gravity of the gas.  As a
consequence, the gas disk's internal 
dynamical time $t_{\rm dyn}\sim H/\sigma_z$ (or the
timescale for vertical oscillations $\sim 2 \pi H/\sigma_z$) 
varies as $t_{\rm dyn}\propto (G \rho_*)^{-1/2}$ rather than
$t_{\rm dyn}\propto \tff(\rho_0) \propto (G \rho_{0})^{-1/2}$,
yielding an outer-disk star formation rate $\sigsfr \propto \Sigma 
\sqrt{\rho_*}$ (see equations [22] and [A16] in \citet{Ostriker2010}).

For turbulent, molecular-gas-dominated starburst regions where
equation (\ref{SFR-tff-eq}) is hypothesized 
to apply, the value of
$\epsff(\rho_0)$ is uncertain, but analysis of observations including 
nearby star-forming clouds, and both moderate-density and high-density 
extragalactic molecular gas, is consistent with a 
range $\epsff(\rho_0)\sim 0.003-0.06$ \citep{KT07,Bigiel08,2009ApJS..181..321E,
2009ApJ...704..842B}.  Simulations of turbulent, 
star-forming disks can also be used to evaluate $\epsff(\rho_0)$.  The mean 
midplane density $\rho_0$ can be measured and used to obtain a mean value of 
$\tff(\rho_0)$; if the value of $\sigsfr$ measured in the simulations is 
multiplied by the measured $\tff(\rho_0)$ and divided by $\Sigma$, the 
result is a numerical evaluation of $\epsff(\rho_0)$.  For the numerical 
simulations described in Section 3, the models yield a range 
 $\epsff(\rho_0)\sim 0.004-0.01$ 
(for further discussion, see Shetty \& Ostriker 2010, in
preparation).  As a fiducial value, we shall adopt $\epsff(\rho_0)=0.005$, but 
it is important to keep in mind that further observations and simulations are 
needed to determine 
this value (and any dependence on parameters) more conclusively.

A disk that is in vertical equilibrium will obey equation 
(\ref{Pmid_eq}); assuming gas thermal+turbulent 
pressure exceeds other forms of support (${\cal R}\ll 1$), 
$\rho_0= \pi G \Sigma^2(1 + \chi)/(2 \sigma_z^2)$.  Substituting this expression
for $\rho_0$ in equation (\ref{SFR-tff-eq}), 
the star formation rate can be written as:
\begin{eqnarray}
\sigsfr
&=&   \frac{4 \epsff(\rho_0) (1+\chi)^{1/2}}{\sqrt{3}}
\frac{G   \Sigma^2}{\sigma_z}
\nonumber \\
&=& 0.051 \Msun \kpc^{-2} \yr^{-1}
(1+\chi)^{1/2}
\left(\frac{\epsff(\rho_0)}{0.005}\right) 
\left(\frac{\sigma_z }{10 \kms} \right)^{-1}
\left(\frac{\Sigma}{100 \Msun\pc^{-2}} \right)^{2}.
\nonumber
\\
\label{Sigma_SFR_alt}
\end{eqnarray}
Thus, disks dominated by self- rather than external-gravity ($\chi \ll 1$)
will have $\sigsfr \propto \Sigma^2$ if 
$\epsff(\rho_0) \sim constant$ and 
the velocity dispersion 
varies only weakly with $\Sigma$  -- as we verify below.

We now focus on the case in which turbulence driven by star formation 
dominates the 
vertical support and control of star formation.  In this limit,
equating (\ref{Sigma_SFR_turb}) with (\ref{Sigma_SFR_alt}) 
for $\sigma_z \rightarrow v_z$ 
yields: 
\begin{eqnarray}
v_z&=& \frac{2 f_P \epsff(\rho_0) p_*  }{ \sqrt{3} \pi m_*}
\frac{1}{\left(1+ \chi \right)^{1/2}  }
\nonumber\\
&=&5.5\kms
\frac{f_p}{\left(1+ \chi \right)^{1/2}  }
\left(\frac{\epsff(\rho_0) }{0.005 } \right)
\left(\frac{p_*/m_*}{3000 \kms } \right).
\label{v_z_eq}
\end{eqnarray}
The only dependence of the velocity dispersion on the mean gaseous surface
density $\Sigma$ is through $\chi$; if the disk is marginally unstable
($Q\simlt 2$), however, $\chi \simlt 0.5$ (see Section 2).  In the
supernova-driven, turbulence-dominated limit, the velocity dispersion
is therefore expected to vary only weakly with $\Sigma$ for disks that
are marginally gravitationally unstable.  The numerical simulations
described in Section 3 indeed show a very weak dependence of $v_z$ on $\Sigma$,
with turbulent velocity dispersions varying by a factor two ($\sim 4-9
\kms$) in models with gas surface density between $\Sigma=100 \Msun
\pc^{-2}$ and $\Sigma =800 \Msun \pc^{-2}$, which have $\sigsfr$
varying by more than a factor 100.  
The main parameter controlling the velocity dispersion, in disks that cool
strongly and therefore have highly-supersonic turbulence, is the specific 
momentum input rate associated with star formation, $p_*/m_*$.
In equation (\ref{v_z_eq}), the fiducial
value of $p_*/m_*$ chosen is that associated with cooled supernova shells.  In
addition to the direct momentum input from radiative blast waves,
cosmic rays that are accelerated by blast waves may drive additional
turbulence in escaping from the disk (e.g. via Parker instabilities),
which would raise the value of $p_*$ and increase $v_z$.

By combining equations (\ref{v_z_eq}) and (\ref{chi_def}) (for ${\cal
  R}\ll 1$), we can solve for $\chi$ and
$C=\chi(1+\chi)$ in terms of the 
star formation parameters and the gas disk and bulge properties. Defining
\begin{eqnarray}
{\cal A} &\equiv& 
\frac{2\zeta_d  }{\pi }
\left(\frac{4 f_P \epsff(\rho_0) p_*}{3 \pi  m_* }  \right)^2  
\frac{\rho_b }{G \Sigma^2   }
\label{A_def}\\
&=&0.11
\left(\frac{\epsff(\rho_0)  }{0.005  }  \right)^2
\left(\frac{p_*/m_* }{3000 \kms  }  \right)^2
\left(\frac{\Omega  }{0.1 \Myr^{-1}  }  \right)^2
\left(\frac{\Sigma  }{100\Msun \pc^{-2}  }\right)^{-2}
\end{eqnarray}
we obtain the implicit relation $\chi(1+\chi)^2={\cal A}$.  
An approximate solution good to within 
15\% over all $\cal A$ is 
\begin{equation}
\chi\approx \frac{{\cal A}}{1+{\cal A}^{2/3}}.
\label{chi_ap_eq}
\end{equation}
For ${\cal A} \ll 1$, $\chi \approx C \approx {\cal A}$, and 
the Toomre parameter is $Q = 2.5 (\sigma_R/\sigma_z) {\cal A}^{1/2}$.

In the turbulence-dominated limit $\sigma_z \approx v_z$ 
(and for ${\cal  R}\ll 1$), 
the mean midplane density 
$\rho_0\equiv n_0 (1.4 m_H)$ (for $m_H$ the hydrogen mass)  
and half-thickness $H\equiv \Sigma/(2\rho_0)$ are given by:
\begin{equation}
\rho_0=\frac{(1+ \chi)\pi}{2}\frac{ G \Sigma^2}{v_z^2}
=\frac{3\pi  }{2  }(1+\chi)^2 
\left(\frac{ \pi  m_*}{2 f_p \epsff(\rho_0) p_*  }  \right)^2 
G\Sigma^2
\label{rho_0_eq}
\end{equation}
so that 
\begin{equation}
n_0=64 \cm^{-3}
\frac{\left(1+ \chi   \right)^2}{f_p^2} 
\left(\frac{\epsff(\rho_0) }{0.005 } \right)^{-2}
\left(\frac{p_*/m_* }{3000\kms} \right)^{-2}
\left(\frac{\Sigma }{100\Msun\pc^{-2} } \right)^2,
\label{density-eq}
\end{equation}
and 
\begin{eqnarray}
H&=&\frac{1}{(1+\chi) }\frac{ v_z^2 }{\pi G \Sigma^2 }     
=\frac{1}{3\pi }
\frac{1}{(1+\chi)^2}
\left(\frac{2 f_p\epsff(\rho_0) p_*}{\pi m_*}\right)^2
\frac{1}{G\Sigma}
\nonumber\\
&=&23\pc
\frac{ f_p^{2} }{\left(1+ \chi   \right)^{2}} 
\left(\frac{\epsff(\rho_0) }{0.005 } \right)^{2}
\left(\frac{p_*/m_* }{3000\kms} \right)^{2}
\left(\frac{\Sigma }{100\Msun\pc^{-2} } \right)^{-1}.
\label{H_eq}
\end{eqnarray}

We have written equations (\ref{Sigma_SFR_turb}), (\ref{v_z_eq}),
(\ref{rho_0_eq}), and
(\ref{H_eq}) in such a way as to highlight the
dependence on the gas surface density.  As $\Sigma$ increases, and
taking ${\cal A}, \chi \ll 1$, the surface density of star formation increases as
$\sigsfr \propto \Sigma^2 m_*/p_*$, the vertical velocity
dispersion is constant $v_z \propto \epsff(\rho_0) p_*/m_*$, the mean
midplane gas density increases as $\rho_0 \propto \Sigma^2 (\epsff
(\rho_0)p_*/m_*)^{-2}$ and the scale height decreases as $H \propto
\Sigma^{-1} (\epsff(\rho_0) p_*/m_*)^2$.  Increasing the specific momentum
input $p_*/m_*$ from star formation decreases the star formation rate
and increases the velocity dispersion and disk thickness.  
The  absence of any explicit timescale in equation (\ref{Sigma_SFR_turb})
is a signature of self-regulation: 
the star formation rate must adjust until the turbulence driven by feedback 
from young stars provides a pressure that matches the vertical weight of
the gas.  If the turbulence level were lower than the equilibrium value, the 
mean density would be higher than equilibrium ($\rho \propto v_z^{-2}$), 
which would then lead to a 
shorter dynamical time $t_{ff}\propto \rho^{-1/2}$ and consequently (from 
equations [\ref{SFR-tff-eq}] and [\ref{Sigma_SFR_alt}]) a higher
star formation rate, driving the velocity dispersion upwards.  Similarly,
too high a turbulence level would lead to lower-than-equilibrium 
$\rho_0$ and $\sigsfr$, which would reduce turbulent driving 
from star formation and eventually
lead to a lower value of $v_z$.  

Although a balance between turbulent driving and dissipation has not
been explicitly used in order to derive the above results, it is
straightforward to see that these considerations yield
equivalent results.  The turbulent vertical momentum per unit time
per unit gas mass that is driven by stellar inputs is 
$\sim p_* \sigsfr/(\Sigma m_*)$.  Since the dissipation time for turbulence 
is comparable to the flow crossing time over the largest scale 
\citep{Stone98,MacLow98}, which in
this case is the disk thickness $H\sim v_z^2/(G \Sigma)$, the 
dissipation rate of momentum per unit gas mass is 
$\sim v_z^2/H \sim G \Sigma$.  Equating driving with dissipation yields
$\sigsfr \sim G \Sigma^2 m_*/p_*$, which is the same as equation 
(\ref{Sigma_SFR_turb}) up to order-unity dimensionless constants 
and the factor $f_p$ that has been incorporated to parameterize the 
details of the momentum injection and mixing.  

In terms of the feedback and disk parameters, the Toomre parameter
for turbulence-dominated disks is
\begin{eqnarray}
Q=0.8\frac{ 1 }{(1+\chi)^{1/2}   }
\frac{ \sigma_R }{\sigma_z   }
\left(\frac{\epsff(\rho_0)  }{0.005  }  \right)
\left(\frac{p_*/m_* }{3000 \kms  }  \right)
\left(\frac{\Omega  }{0.1 \Myr^{-1}  }  \right)
\left(\frac{\Sigma  }{100\Msun \pc^{-2}  }\right)^{-1}.
\end{eqnarray}
Thus, for $\Sigma$ and $\Omega$ in the range observed for galactic center 
starbursts, the disk will be near the margin of instability.  Even if 
$Q\simgt 2$, the intermittency of turbulence implies that some regions 
may temporarily become unstable.  If a large region of the disk becomes
unstable and begins to contract, a gravitationally bound supercloud may
form.

In a gravitationally bound supercloud, the outward force $F_{\rm kin}$
from direct kinetic feedback from star formation varies as $\sim \dot
M_* p_*/m_*$, whereas the outward force $F_{\rm rad}$ from 
reprocessed radiation
varies as $\sim L_* \tau/c \propto \dot M_* M/r^2$, where $\dot M_*$,
$L_*$, $M$, and $r$ are the supercloud's star formation rate,
luminosity, mass, and radius.  The inward gravitational force $F_{\rm
  grav}$ varies as $\sim G M^2/r^2$.  If $\dot M_* \propto
M/\tff(\bar\rho)\propto r^{-3/2}$, then $F_{\rm kin}/F_{\rm grav}
\propto r^{1/2}$ (decreasing as a supercloud shrinks), whereas $F_{\rm
  rad}/F_{\rm grav} \propto r^{-3/2}$ (increasing as a supercloud
shrinks).  This suggests that superclouds, if they become bound,
could be destroyed by radiation pressure feedback but not by kinetic
feedback.  In sufficiently optically thick disks, radiation pressure
might prevent supercloud formation; a simple estimate in Appendix B
suggests that disks with gas surface densities up to $\Sigma \sim
(3000/\kappa_{\rm IR}) \Msun \pc^{-2}$ may be susceptible to
supercloud formation.

\section{Summary and discussion}

Increasingly high-resolution, multiwavelength observations of galaxies
have refined empirical relations between the gas content and mean star
formation rate in galaxies.  The picture now emerging appears to
include at least three regimes: outer-disks, mid-disks, and galactic
centers.  The first and last regimes evidence superlinear mean
relations between the surface density of star formation $\sigsfr$ and
the gas surface density $\Sigma$, while the relation is close to
linear for mid-disks.

From a theoretical point of view, star formation in all regimes is
likely to be self-regulated in some way. For the outer-disk regime,
\citet{Ostriker2010} have proposed that UV feedback from massive young
stars is particularly important to self-regulation.  In this model,
the star formation rate in outer disks must be such that the
heating of diffuse atomic gas by stellar UV radiation balances cooling,
with the cooling rate set by the gas pressure (and hence responsive to
the vertical gravity of the disk).  For the mid-disk regime, gas
resides predominantly in gravitationally-bound (but transient) GMCs
with mean densities (and hence star formation rates) that depend on
the amplitude of supersonic turbulence driven by star formation
feedback (possibly by expanding HII regions, although this is not well
understood).  For galactic center regions, we argue in this paper that
the momentum injected to the disk by massive star formation is crucial
to establishing a self-regulated state.  Vertical equilibrium of the
disk requires gravity to be balanced by an outward momentum flux,
consisting primarily of turbulent pressure (largely driven by
supernovae) for most molecule-dominated central regions.  Our analysis
suggests that the galactic-center star formation regime may further
subdivide, with a transition to radiation-dominated momentum flux (or
pressure) only for extremely high $\Sigma$, highly-opaque disks.

The analysis of Section 2 provides a prediction (see equation
\ref{Sigma_SFR_P_eq}) for the dependence of $\sigsfr$ on $\Sigma$ and
the density $\rho_b$ of the stellar bulge, parameterized by the mean
specific momentum $p_*/m_*$ injected by star formation to the ISM
(primarily from supernovae), by the mean IR opacity $\kappa_{\rm IR}$
(which determines how much reprocessed starlight is trapped), and by
the mass-to-radiation energy efficiency of star formation $\epsilon_*$
(which depends on the IMF).  Assuming that gas gravity dominates the
vertical potential, self-regulated, turbulence-dominated disks
are expected to have $\sigsfr \approx 2 \pi m_* G \Sigma^2/p_*$ (see
equation \ref{Sigma_SFR_turb}), whereas radiation-dominated disks are
expected to have $\sigsfr \approx 2 \pi G \Sigma/(\epsilon_*
 c \kappa_{\rm IR})$ (see equation \ref{Sigma_SFR_rad}).  The
turbulence-dominated regime is expected to apply for $\Sigma \sim
100-10^4 \Msun \pc^{-2}$, covering most starbursts (although not the
innermost disk regions that merge into AGN disks).  For the
radiation-dominated limit, an equivalent expression to our result was
previously obtained (under somewhat different assumptions) by
\citet{TQM05} (see their equation 28), and they also noted that
optically-thin disks are expected to have a steeper dependence of
$\sigsfr$ on $\Sigma$ than optically-thick disks.

In a self-regulated disk, the feedback parameters determine the star
formation rate, independent of the internal structure of the disk.
Thus, the vertically-integrated star formation rates of equations
(\ref{Sigma_SFR_turb}) and (\ref{Sigma_SFR_rad}) depend only on the
vertically-integrated disk properties.  However, it is natural that
the star formation rate should also connect to the timescale for gas
to become concentrated into high-density clouds, as discussed in
Section 5.  For a system in equilibrium where the gas is confined
primarily by its own gravitational potential, this timescale is the
gravitational free-fall time at the mean density; for a disk,
$1/\tff(\bar\rho)\propto (G \Sigma/H)^{1/2}$. The mean density (or
disk thickness), in turn, depends on what is balancing gravity --
thermal pressure, turbulent pressure, or radiation pressure.  In the
turbulence-dominated, gas-gravity dominated case, $H= v_z^2/(\pi
G\Sigma)$ for $v_z$ the vertical velocity dispersion, which yields
$1/\tff(\bar\rho)= 4G \Sigma/(\sqrt{3}v_z)$.  If $\epsff(\rho_0)$ is
the collapse efficiency per free-fall time at the mean midplane
density, the star formation rate is given by $\sigsfr =
2.3\epsff(\rho_0) G \Sigma^2/v_z$ (see equation \ref{Sigma_SFR_alt}).

\citet{KM05} proposed that the log-normal density distribution in
highly turbulent systems will lead to a star-forming efficiency
declining as the $-1/3$ power of the turbulent Mach number.  Including
this scaling would lead to $\sigsfr \propto \Sigma^2
v_z^{-1.3}$. Under the assumption that in molecule-dominated gas 
$\sigsfr =\epsff(\rho) \Sigma/\tff(\rho)$, \citet{2009ApJ...699..850K}
proposed a power-law relation between $\sigsfr$ and $\Sigma$ in
galactic center regions.  In obtaining this 
relation, they assumed that the Toomre parameter $Q\approx 1$, and also that
$\Omega \propto \Sigma^{0.5}$; together these would imply that
velocity dispersion varies $\propto \Sigma^{0.5}$.  Although we in
fact argue that the vertical velocity dispersion $v_z$ is rather insensitive to 
$\Sigma$, inserting $v_z \propto \Sigma^{0.5}$ in $\sigsfr \propto
\Sigma^2 v_z^{-1.3}$ would yield $\sigsfr \propto
\Sigma^{1.3}$, the result given in equation (10) of
\citet{2009ApJ...699..850K}.

In order for the disk to form stars at a rate controlled by the
free-fall time ($\sigsfr \approx 2.3\epsff(\rho_0) G \Sigma^2/v_z$), and
also for the momentum feedback from star formation to control vertical
equilibrium of the disk ($\sigsfr \approx 2 \pi m_* G \Sigma^2/p_*$),
the vertical velocity dispersion must adjust 
to $v_z\approx 0.4 \epsff(\rho_0) p_*/m_*$
(see equation \ref{v_z_eq}).  Interestingly, this is independent of
the disk surface density and the star formation rate; our numerical
simulations in fact show that $v_z$ is insensitive to the star
formation rate.  Other recent numerical simulations have also
indicated that turbulent velocity dispersions vary relatively weakly
with the input star formation rate \citep{ 2006ApJ...638..797D,
  2008ApJ...684..978S, 2009MNRAS.392..294A, 2009ApJ...704..137J}.  The
relation $v_z \sim \epsff(\rho_0) p_*/m_*$ can also be understood as
simply a balance between driving turbulent velocities in the ISM at a
rate $\sim p_* \epsff(\rho_0)/[m_*\tff(\rho_0)]$, and
dissipating turbulent velocities at a rate $\sim v_z  v_z/H \sim
v_z/\tff(\rho_0)$.  For $\epsff(\rho_0)\sim 0.01$ and $p_*/m_*\sim 3000 \kms$, 
the velocity dispersion driven by star formation feedback is $v_z\sim 10\kms$.  

More generally, if the vertical momentum injection rate per unit ISM gas mass 
from star formation is $\sim (1-2)p_*/(4 m_* \tsfgas)$, and the 
corresponding turbulent dissipation rate is $\sim v_z/[(1-2)t_{\rm dyn}]$ 
for vertical dynamical time $t_{\rm dyn} = H/\sigma_z$,
the expected vertical velocity dispersion is 
$v_z \sim (0.3-1) (p_*/m_*) t_{\rm dyn}/\tsfgas$.  This is
within a factor of a few of $10\kms$ if 
 $p_*/m_* \sim 3000 \kms$ from supernovae and 
$\tsfgas \sim 100 t_{\rm dyn}$ (which is true in outer disks as well as 
starbursts; see below),
regardless of whether the vertical 
dynamical time is set by gas self-gravity or by gravity of an external potential
(stellar bulge, stellar disk, or dark matter halo).
Velocity dispersions reported for  
starburst regions are often much larger than 
the values we find (e.g. 
\citealt{1998ApJ...507..615D,2010arXiv1011.5360G}),
 but it is difficult to eliminate 
sub-beam sheared rotation, radial/azimuthal streaming, 
and other non-turbulent (and non-vertical) motions 
on the relevant scales ($\sim H \ll 100 \pc$).  Observations 
of molecular velocity dispersions within the central kpc of a 
few nearby, face-on galaxies are $\sim 10\kms$ \citep{1997A&A...326..554C}; 
turbulent molecular velocity dispersions from larger
scales are also similar \citep{2010MNRAS.tmp.1674W}.

The momentum-regulated, turbulence-dominated prediction of equation
($\ref{Sigma_SFR_turb}$) for $\sigsfr$ can be compared to the results
of numerical simulations and to observations.  We do this in Sections 3
and 4, respectively.  The simulations indeed yield a relation
consistent with the prediction, $\sigsfr \approx 0.1 \Msun \kpc^{-2}
\yr^{-1} (\Sigma/100 \Msun \pc^{-2})^2$, adopting a value of the
momentum input per unit stellar mass formed of $p_*/m_*=3000 \kms$.  Velocity 
dispersions from the simulations are also comparable to prediction.
Further comparisons of the star formation rates and disk properties 
from the model with the results of numerical simulations will be
presented in Shetty \& Ostriker (2010, in preparation).  

A difficulty in comparing to observations is that the molecular mass
must be obtained indirectly through observations of CO transitions.
Although the conversion factor $X_{\rm CO}$ from integrated CO
intensity to hydrogen column density has been well calibrated
empirically using several different methods for individual GMCs and
main-disk regions which integrate over collections of GMCs, the value
of $X_{\rm CO}$ is much less certain in galactic center regions.
There is evidence that $X_{\rm CO}$ decreases in high-$\Sigma$
starburst regions; this may owe to a combination of factors, including
higher CO excitation in the warmer, denser gas, a larger
fraction of the gas that is sufficiently shielded to create CO, and
larger velocity gradients that allow radiation to escape more easily.
We compare to the large sample of observations compiled in 
\citet{2010MNRAS.407.2091G}.
If the conversion factor is assumed to decrease with increasing
integrated CO intensity (as $X_{\rm CO} \propto I_{\rm CO}^{-0.3}$),
the observed and predicted relations for $\sigsfr$ vs. $\Sigma$ are in
excellent agreement, for the range $\Sigma \sim 10^2-10^4 \Msun
\pc^{-2}$.  If two different constant conversion factors are adopted
for normal and merger systems, the former agree with the predicted
$\sigsfr$ at low $\Sigma$, and the latter agree at high $\Sigma$.  To
make progress in relating theory to observations, 
a top priority is to determine how $X_{CO}$ varies under the conditions that 
prevail in galactic center regions. 
This will require both empirical calibrations, and
radiative transfer models applied to numerical simulations that
resolve the structure of turbulent disks.

Finally, we note that the star formation timescale $\tsfgas =
\Sigma/\sigsfr \approx p_*/(2 \pi G m_* \Sigma)$ in self-regulated,
turbulence-dominated, molecular galactic center disks can also be
expressed as $\tsfgas\sim t_{\rm dyn}/\epsff(\rho_0)\sim 100 t_{\rm
  dyn}$, for vertical dynamical time $t_{\rm dyn} = H/\sigma_z$.  In
\citet{Ostriker2010}, we showed\footnote{See equation (A16) in
  \citet{Ostriker2010}; note that $t_{\rm con}$ there is the same as
  $\tsfgas$ here.} that in outer disks, self-regulation via UV heating
of atomic gas leads to a quantitatively similar result, $\tsfgas \sim 200
H/\sigma_z$.  The constant factor $\sim 200$ 
in the outer-disk formula for $\tsfgas$
is set, however, by the ratio of heating and cooling rate coefficients
for atomic gas and by the UV production efficiency of the young-star
population, rather than by the turbulent processes that set the
constant factor $\sim 1/\epsff(\rho_0)$ in the galactic-center formula.  Thus,
although very different physical processes control star formation in
galactic centers and outer disks, the dependence of $\sigsfr$ on
$1/t_{\rm dyn}$ (and on $\Omega \sim Q/t_{\rm dyn}$) is similar, both
in terms of the scaling and (coincidentally) the coefficient.  We
believe that this dependence on $1/t_{\rm dyn}$, together with the
limited range of $Q$ that disk evolution yields, is the reason that
both whole galaxies and central starbursts, at a range of redshifts, tend to
follow the empirical relation $\sigsfr \sim 0.02 \Sigma \Omega$
\citep{Kennicutt98,2010ApJ...714L.118D,2010MNRAS.407.2091G}.

\acknowledgements We are grateful to Reinhard Genzel for providing a
table based on his compilation of galactic gas content and star
formation rates.  We thank the referee for helpful suggestions on 
the manuscript.
The work of E.C.O. was supported by a fellowship
from the John Simon Guggenheim Foundation, and by grant AST-0908185
from the National Science Foundation.  
R.S. acknowledges support from the German Bundesministerium f\"ur Bildung
und Forschung via the ASTRONET project STAR FORMAT (grant 05A09VHA).

\appendix

\section{Appendix: momentum injection to the ISM 
from radiation-driven shells around star clusters}

Consider a GMC of mass $\mgmc$ which forms stars with a net
efficiency over its lifetime of $\epsgmc\equiv M_*/\mgmc$, ejecting
the remaining gas mass $(1-\epsgmc)\mgmc$.  Treating the stellar
component as collected in a single central cluster of (fixed) radius
$r_{cl}$ and luminosity $L_*=\epsgmc \Psi \mgmc$, and the ejected mass
as an expanding spherical shell of (variable) radius $r$ surrounding
it, the total radiation force applied to the interior of the shell 
by reprocessed (diffuse) radiation is 
$F_{\rm  rad}=4\pi r^2 (a_{\rm rad}/3)T_{\rm int}^4$, where 
$T_{\rm  int}^4=(3/4)\tau T_{\rm eff}^4$ for shell optical depth $\tau\gg 1$
and internal and surface (effective) temperatures $T_{\rm int}$ and
$T_{\rm eff}$. Equating $L_*$ with 
$4\pi r^2 (a_{\rm rad} c/4) T_{\rm  eff}^4$ and using 
$\tau=\kappa_{\rm IR} (1-\epsgmc) \mgmc /(4\pi r^2)$, 
$F_{\rm rad} =L_* \tau/c = \kappa_{\rm IR} (1-\epsgmc)\epsgmc
\Psi\mgmc^2 /(4\pi c r^2)$.  The gravitational force on the shell,
including that of the central cluster and its own self-gravity, is 
$-G\mgmc^2(1-\epsgmc^2)/(2 r^2)$.  The net acceleration of the shell is
therefore
\begin{equation}
\ddot r = \frac{G\mgmc}{r^2}\left[\frac{\kappa_{\rm IR}\Psi \epsgmc }{4\pi G c } 
-\frac{(1+\epsgmc)  }{ 2 } \right]\equiv 
\frac{(\psi\epsgmc -1) }{2}\frac{ G\mgmc}{r^2} 
\label{cluster_eom}
\end{equation}
for $\psi \equiv \Psi \kappa_{\rm IR}/(2\pi c G)-1$.
As pointed out by \citet{2010ApJ...709..191M}, this implies that there is a 
minimum value of $\epsgmc$ for the pressure associated with reprocessed 
radiation to disrupt the cloud; the shell can only become unbound if 
$\epsgmc > \psi^{-1}\equiv \varepsilon_{min} $. 

Equation (\ref{cluster_eom}) can be integrated 
to obtain the asymptotic speed $v_f = [(\psi\epsgmc -1)G\mgmc /r_{cl}]^{1/2}$ 
of the shell,
assuming its initial radius is $r_{cl}$.  The resulting ratio of the momentum
in the shell divided by the total mass in stars formed, representing 
$p_*/m_*$ for a radiation-driven shell, is $v_f (1-\epsgmc)/\epsgmc$.  For
any given value of 
$\psi$, it is straightforward 
to show that the
maximum momentum-to-mass ratio is obtained when 
$\epsgmc = 4 \varepsilon_{min} [1+ ( 1 + 8 \varepsilon_{min})^{1/2}]^{-1} 
\equiv \varepsilon_{max}$.
Using 
$\Psi =2000 \; {\rm erg}\;{\rm s}^{-1}\g^{-1}$
\citep{2006ApJ...647..244D}
for young, luminous clusters 
that fully sample the \citet{Kroupa01} IMF 
and $\kappa_{\rm IR}=20 \cm^2 \g^{-1}$ 
for warm, dusty shells \citep{2010ApJ...709..191M,2003A&A...410..611S},
this yields $\varepsilon_{min}= 0.5$, $\varepsilon_{max}=0.6$, and 
$[p_*/m_*]_{\rm rad}= 0.4( G \mgmc/r_{cl})^{1/2}= 
30 \kms (\mgmc/10^6\Msun)^{1/2}(r_{cl}/\pc)^{-1/2} $.  
This is comparable to the
velocity dispersion of the cluster itself, 
which even for the most massive clusters
in starbursts does not exceed $ 100\kms$ and is typically much lower
\citep{2003ApJ...596..240M,2004A&A...416..467M,2009ApJ...706..203O}.  

In addition to diffuse (reprocessed) radiation, streaming radiation
(dominated by UV) from the star cluster also imparts momentum to the 
expanding gas cloud (or shell)
where it is first absorbed.  The maximum force from this streaming
radiation is $L_*/c$, such that the maximum contribution to $v_f
(1-\epsgmc)/\epsgmc=p_*/m_*$ is $\int \Psi dt/c = \epsilon_* c \sim
200 \kms$. Thus, allowing for both reprocessed and direct stellar
radiation, the value of $p_*/m_*$ from radiation-driven expanding
shells is smaller than the momentum-per-stellar mass injected to the
disk by radiative supernovae shells, $\sim 3000 \kms$.

\section{Appendix: supercloud evolution}

Potentially, large-scale instabilities in the disk can lead to gaseous
supercloud formation.  How would a supercloud evolve if it is able to
form?  For a spherical, uniform-density cloud of mass $M$ and radius
$r$ that forms stars at a rate $\dot M_*=\epsff(\bar\rho) M/\tff(\bar
\rho)$, producing a luminosity $L_*= \epsilon_* c^2 \dot M_*$ and a
supernova rate $\dot M_*/m_*$, the ratio of the outward force (due to
the combined momentum input from supernovae and reprocessed 
radiation pressure\footnote{Streaming radiation adds a term 
$< \epsilon_* c \sim 200 \kms $ to $p_*/m_*$ that is small compared 
to the contribution $p_*/m_* \approx 
3000 \kms$ from supernovae, and thus may be neglected.}) to
the inward force (due to gravity) is
\begin{equation}
\frac{F_+}{F_-}=\frac{2^{7/2} \epsff(\bar\rho)  }{3 \pi G^{1/2} M^{1/2}}
\left[ f_p \frac{ p_*}{ m_*} r^{1/2} + 
\frac{3 \epsilon_* c \kappa_{\rm IR} M }{4\pi   } r^{-3/2}\right].
\end{equation}
The contribution to $F_+/F_-$ from supernovae declines $\propto
r^{1/2}$ as $r$ decreases, whereas the contribution to $F_+/F_-$ from
radiation increases $\propto r^{-3/2}$.  Thus, radiation forces become
increasingly important as the cloud shrinks; eventually, if $F_+/F_-$
becomes $>1$, inward contraction will be halted. If the star formation
efficiency is sufficient, the cloud will ultimately be destroyed, with
a substantial fraction of its mass once again becoming unbound
(cf. \citealt{2009ApJ...703.1352K,2010ApJ...709..191M,2010MNRAS.tmp..635K}
and the discussion in Appendix A).  

The masses of clouds that form
via gravitational instability in disks with $Q< Q_{\rm crit}$ are
typically $\sim 1-10 M_{\rm J, 2D}$
(e.g. \citealt{2002ApJ...581.1080K,2003ApJ...599.1157K,2007ApJ...660.1232K}), where $M_{\rm J,2D} = (\pi H)^2 \Sigma = \sigma^4/(G^2\Sigma)$ for
$\sigma$ the gas velocity dispersion.  Letting $M\equiv m M_{\rm
  J,2D}$, $r\equiv x H$, and assuming $\sigma=v_z$, we can use
equations (\ref{v_z_eq}), (\ref{H_eq}), (\ref{tau*_eq}), and
(\ref{tau_eq}) to obtain
\begin{equation}
\frac{F_+ }{F_-} = \frac{1.8}{m^{1/2}}
\left[x^{1/2} +  \frac{3\pi}{4}\frac{\tau}{\tau_*} \frac{m}{x^{3/2}} \right],
\end{equation}
where $\tau$ is evaluated using the unperturbed disk surface density $\Sigma$,
and we have assumed $\chi \ll 1$.  

For a given $m$, the force ratio has a minimum at $x_{\rm min}=2.7(m
\tau/\tau_*)^{1/2}$, where its value is $(F_+/F_-)_{\rm min} = 4.0
m^{-1/4} (\tau/\tau_*)^{1/4}$.  Taken at face value, this suggests
that only disks with a sufficiently low surface density $\Sigma$ will
form contracting superclouds, since only if $\tau/\tau_*$ is
sufficiently small can $F_+/F_-$ be $<1$.  With $m= 10$, $F_+/F_-<1$
for $\tau/\tau_* <0.04$, corresponding to $\Sigma <(3000/\kappa_{\rm
  IR}) \Msun \pc^{-2} $.  Note also that $x_{\rm min}<1$ only for 
$\Sigma <(1000/\kappa_{\rm  IR}) \Msun \pc^{-2} $.
Of course, given the highly idealized
assumptions adopted, the particular value obtained from this simple
estimate should not be taken too seriously.  Physically, however, it
is reasonable to expect higher-$\Sigma$ 
disks that are increasingly radiation-pressure
supported to be less subject to gravitational instability, because of
their stiff equation of state: if $P_{\rm rad} \propto L_* \rho /r$
and $L_* \propto M \rho^{1/2}$, the internal radiation pressure in a
supercloud would vary as $P \propto \rho^{11/6}$.  Potentially, more
detailed understanding of large-scale gravitational instabilities in
starburst disks could be obtained via an analysis similar to that of
\citet{2008ApJ...684..212T}, but including sources of radiation and
turbulence.


\end{document}